\pdfoutput=1

\documentclass[11pt,twoside,a4paper,cmspaper,final,collab]{cms-tdr}
\def\svnVersion{450757P}\def\cmsCernNoTag{CERN-EP-2017-284}\def\cmsCernDate{\today}\def\cmsMessage{Published in Physics Letters B as \href{http://dx.doi.org/10.1016/j.physletb.2018.02.045}{\doi{10.1016/j.physletb.2018.02.045.}}}

\begin{document}\cmsNoteHeader{SUS-16-046}

\hyphenation{had-ron-i-za-tion}
\hyphenation{cal-or-i-me-ter}
\hyphenation{de-vices}
\RCS$Revision: 444364 $
\RCS$HeadURL: svn+ssh://svn.cern.ch/reps/tdr2/papers/SUS-16-046/trunk/SUS-16-046.tex $
\RCS$Id: SUS-16-046.tex 444364 2018-02-05 13:37:07Z jschulz $
\newlength\cmsFigWidth
\ifthenelse{\boolean{cms@external}}{\setlength\cmsFigWidth{0.98\columnwidth}}{\setlength\cmsFigWidth{0.47\textwidth}}
\ifthenelse{\boolean{cms@external}}{\providecommand{\cmsLeft}{top\xspace}}{\providecommand{\cmsLeft}{left\xspace}}
\ifthenelse{\boolean{cms@external}}{\providecommand{\cmsRight}{bottom\xspace}}{\providecommand{\cmsRight}{right\xspace}}
\newcommand{\x}{\ensuremath{\phantom{0}}}
\newcommand{\y}{\ensuremath{\phantom{.}}}
\newcommand{\z}{\ensuremath{\phantom{>}}}

\newcommand{\ST}{\ensuremath{S_{\mathrm{T}}^{\gamma} }\xspace}
\newcommand{\PV}{\ensuremath{\mathrm{V}}\xspace}
\newcommand{\Vgamma}{\ensuremath{\mathrm{V}(\gamma)}\xspace}
\newcommand{\gammaJets}{\ensuremath{\gamma\text{+jets} }\xspace}
\newcommand{\ttgamma}{\ensuremath{\ttbar(+\gamma)}\xspace}
\newcommand{\efake}{\ensuremath{\Pe\to\gamma}\xspace}
\newcommand{\ptm}{\ensuremath{p_{\mathrm{T}}^{\kern1pt\text{miss}}}\xspace}
\newcommand{\mt}{\ensuremath{M_{\mathrm{T}}(\gamma,\ptvecmiss)}\xspace}
\newcommand{\dphi}{\ensuremath{\Delta\phi(\ptvecmiss,\text{nearest jet}/\gamma)}\xspace}

\cmsNoteHeader{SUS-16-046}

\title{Search for gauge-mediated supersymmetry in events with at least one photon and missing transverse momentum in pp collisions at $\sqrt{s} = 13\TeV$}

\date{\today}

\abstract{
A search for gauge-mediated supersymmetry (SUSY) in final states with photons and large missing transverse momentum is presented. The data sample of pp collisions at $\sqrt{s}=13\TeV$ was collected with the CMS detector at the CERN LHC and corresponds to an integrated luminosity of 35.9\fbinv. Data are compared with models in which the lightest neutralino has bino- or wino-like components, resulting in decays to photons and gravitinos, where the gravitinos escape detection. The event selection is optimized for both electroweak (EWK) and strong production SUSY scenarios. The observed data are consistent with standard model predictions, and limits are set in the context of a general gauge mediation model in which gaugino masses up to 980\GeV are excluded at 95\% confidence level. Gaugino masses below 780 and 950\GeV are excluded in two simplified models with EWK production of mass-degenerate charginos and neutralinos. Stringent limits are set on simplified models based on gluino and squark pair production, excluding gluino (squark) masses up to 2100 (1750)\GeV depending on the assumptions made for the decay modes and intermediate particle masses. This analysis sets the highest mass limits to date in the studied EWK models, and in the considered strong production models when the mass difference between the gauginos and the squarks or gluinos is small.
}

\hypersetup{%
pdfauthor={CMS Collaboration},%
pdftitle={Search for gauge-mediated supersymmetry in events with at least one photon and missing transverse momentum in pp collisions at sqrt(s) = 13 TeV},
pdfsubject={CMS},%
pdfkeywords={CMS, physics, software, computing}}

\maketitle

\section{Introduction}
\label{sec:introduction}

The search for physics beyond the standard model (SM) is one of the key research topics of the CMS experiment at the CERN LHC. Especially after the discovery of a Higgs boson with a mass of around 125\GeV in 2012~\cite{Aad:2012tfa, Chatrchyan:2012xdj, Khachatryan:2014jba}, supersymmetry (SUSY)~\cite{Ramond,Ramond:1971kx,Golfand,Volkov,Wess:1974tw,Freedman:1976xh,Deser:1976eh,Freedman:1976py,Ferrara:1976fu,Fayet,Chamseddine,Barbieri,Hall,Kane} is one of the theoretically favored possible extensions of the SM. Among several explanations for unsolved problems in particle physics, SUSY provides a mechanism for stabilizing the SM-like Higgs boson mass at the electroweak (EWK) scale. Since current searches are pushing the limits on strongly produced SUSY particles (sparticles) beyond the one-TeV threshold, the interest in probing gaugino masses via EWK production is growing. While searches for heavy sparticles especially profit from the increase in the center-of-mass energy due to the large increase of the production cross section, searches for EWK production benefit from a larger data set, as collected by the CMS experiment in 2016.

In this Letter, a search for
SUSY focusing on gauge-mediated SUSY breaking (GMSB)~\cite{GGMa,GGMd2,GGMd3,GGMd4,GGMd5,GGMd1,GGMd} scenarios is presented. The $R$-parity~\cite{rparity} is assumed to be conserved, so that SUSY
particles are always produced in pairs. The gravitino (\sGra) is the lightest SUSY particle (LSP)
and escapes undetected, leading to missing transverse momentum (\ptm) in the detector. The
next-to-LSP (NLSP) is assumed to be the lightest neutralino ($\chiz_1$).
Depending on its composition, the $\chiz_1$ can decay according to $\chiz_1\to N\sGra$, where $N$ is
either a photon ($\gamma$), an SM-like Higgs boson ($\PH$), or a $\PZ$ boson. If the
gauginos are nearly mass-degenerate, the chargino ($\chipm_1$) decays $\chipm_1\to
\Wpm\sGra$ are also possible. The \sGra is assumed to have negligible mass and the NLSP is assumed to decay promptly.

The analyzed data set was collected at the CERN LHC in proton-proton collisions at a center-of-mass energy of 13\TeV
and corresponds to an integrated luminosity of 35.9\fbinv. Events are required to contain at least one high-energy photon and large \ptm.
In order to maintain sensitivity to EWK SUSY production, there is no explicit event
selection criterion requiring hadronic energy, i.e., the presence of jets in the event. In GMSB
SUSY, \ptm arises from the stable and noninteracting $\sGra$, while photons originate
from $\chiz_1\to\gamma\sGra$ decays. The energy of the photon as well as of the gravitino and thus
the \ptm is governed by the $\chiz_1$ mass, and the $\chiz_1\to\gamma\sGra$
branching fraction is determined by the neutralino's bino and wino components and its mass.
Compared to analyses requiring photons and large hadronic activity, this analysis has superior sensitivity to GMSB SUSY in EWK production, and also in strong production if the squark, gluino, and the lightest gaugino masses are similar (compressed-spectrum scenarios).

An earlier version of this analysis~\cite{SUS-14-016} was carried out by CMS on a special 8\TeV data set recorded as part of the ``parked-data'' program~\cite{CMS-DP-2012-022} corresponding to an integrated luminosity of
7.4\fbinv using a dedicated trigger and a lower photon transverse momentum (\pt)
threshold of 30\GeV. The ATLAS and CMS collaborations have also searched for direct EWK
production of gauginos in final states with at least one photon and one electron or
muon~\cite{CMS:2015loa,Aad:2015hea}, and in the two-photon
channel~\cite{Chatrchyan:2012bba,Khachatryan:2015exa,Aad:2015hea}. Single-photon and \HT-based
analyses~\cite{Khachatryan:2015exa}, where \HT is the scalar sum of hadronic jet transverse momenta, have good
sensitivity for strong production in GMSB models but lack sensitivity for EWK production and
compressed-spectrum scenarios.

\section{Signal models}
\label{sec:signals}

\begin{figure*}[tbh!]
\centering
\includegraphics[width=\cmsFigWidth]{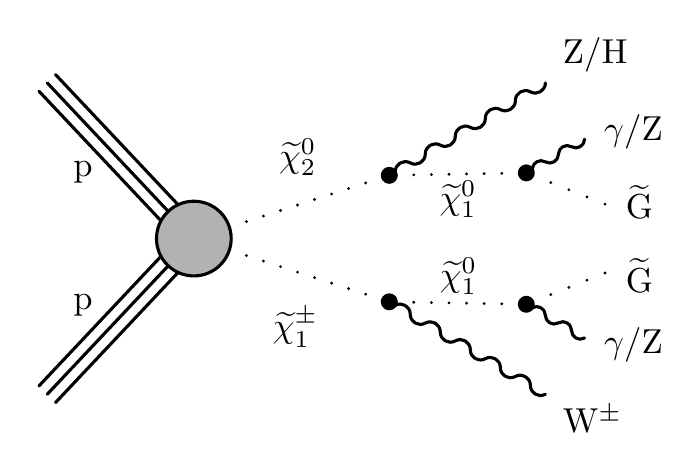}\hfil
\includegraphics[width=\cmsFigWidth]{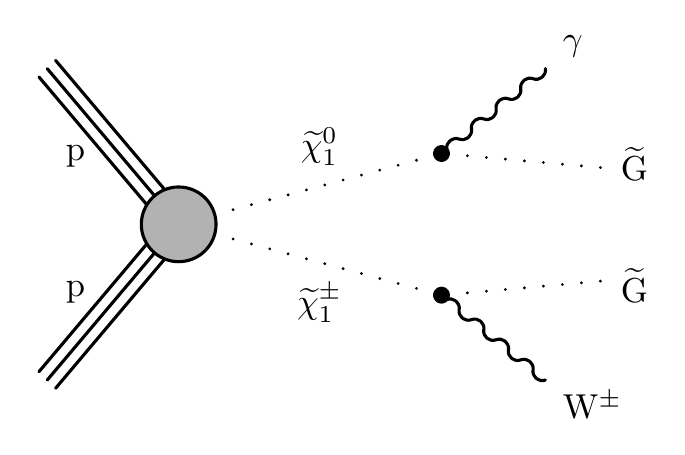}\\
\null
\includegraphics[width=\cmsFigWidth]{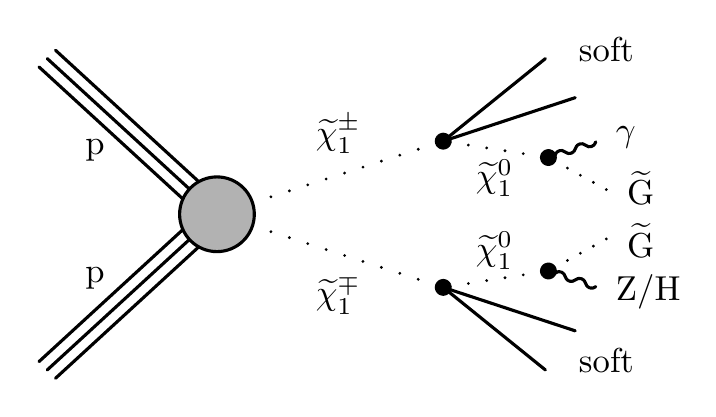}\hfil
\includegraphics[width=\cmsFigWidth]{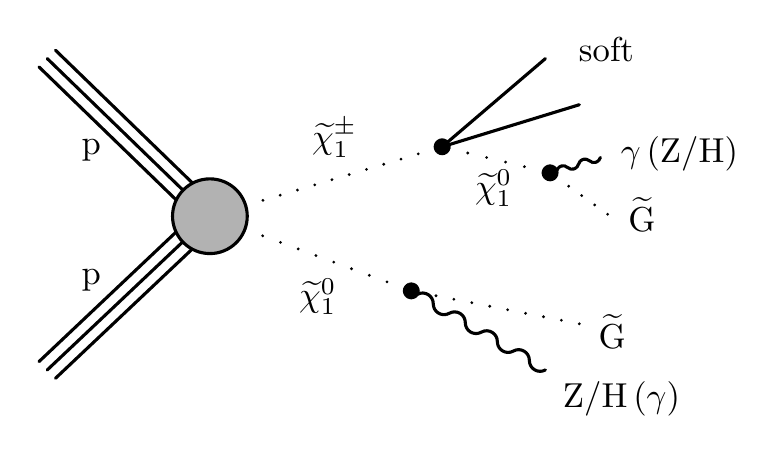}
\caption{
In the context of GGM, several production and decay channels are possible. The diagram of the dominant process $\chiz_2$--$\chipm_1$ production is shown (upper left), where the gaugino decays depend on the mass configuration under study. In the TChiWg model (upper right), the gauginos are mass degenerate. The TChiNg model comprises $\chipm_1$ pair production (lower left) and $\chipm_1\chiz_1$ production (lower right), where the $\chipm_1$ is only slightly heavier than the $\chiz_1$, so only low-momentum (soft) particles appear in the decay of $\chipm_1$ to $\chiz_1$.
\label{fig:GGM}}
\end{figure*}

To interpret the results, a general gauge mediation (GGM)~\cite{GGMe,GGMf,Ruderman:2011vv,Kats:2011qh,Kats:2012ym,Grajek:2013ola} scenario dominated by EWK production is used. Furthermore, two EWK production and four strong production simplified model scenarios (SMS)~\cite{Chatrchyan:2013sza} are considered for interpretation.
For the GGM scenario, the squark and gluino masses are set to a high scale rendering them inaccessible and strong production negligible. The bino and wino masses therefore fully determine the model point under study and are varied in the interpretation. The $\chiz_1$ is assumed to be purely bino-like, while the $\chipm_1$ and $\chiz_2$ are assumed to be purely wino-like. The dominant process for EWK GGM production is shown in Fig.~\ref{fig:GGM} (upper left).
In the GGM framework, where the gauginos are not mass-degenerate by construction, a larger $\chipm$--$\chiz_1$ mass difference increases the hadronic energy in the final state if the $\PZ$, $\PH$, or $\PW$ bosons decay hadronically.

{\tolerance=800
The EWK simplified scenario TChiWg probes associated production of mass-degenerate charginos and neutralinos ($\chipm_1\chiz_1$), assuming the decay modes $\chiz_1\to\gamma\sGra$ and $\chipm_1\to \PW^{\pm}\sGra$, as shown in Fig.~\ref{fig:GGM} (upper right).
The TChiNg scenario assumes nearly mass-degenerate $\chipm_1$ and $\chiz_1$, but considers $\chipm_1\ensuremath{\widetilde{\chi}_{1}^{\mp}}$ and $\chipm_1\chiz_1$ production as shown in Fig.~\ref{fig:GGM} (lower left and right).  In this scenario, the $\chipm_1$ is assumed to have a slightly higher mass than $\chiz_1$, and it decays to $\chiz_1$ and low-momentum particles outside the acceptance of this analysis. The neutralinos are assumed to decay as $\chiz_1\to\gamma\sGra$, $\chiz_1\to \PZ\sGra$, and $\chiz_1\to \PH\sGra$ with 50, 25, and 25\% probability, respectively.
\par}

\begin{figure*}[tbh!]
\centering
\includegraphics[width=\cmsFigWidth]{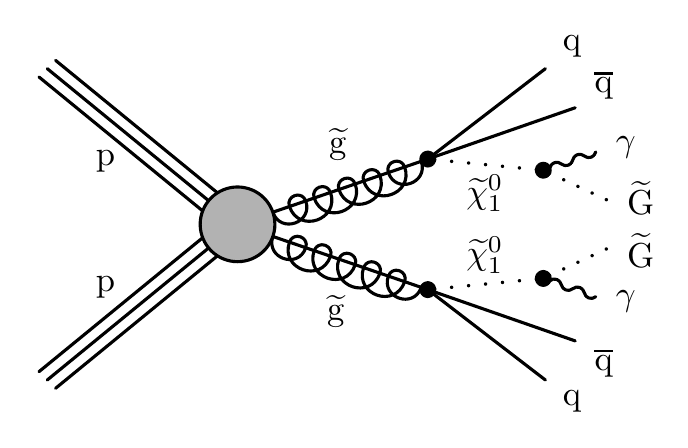}\hfil
\includegraphics[width=\cmsFigWidth]{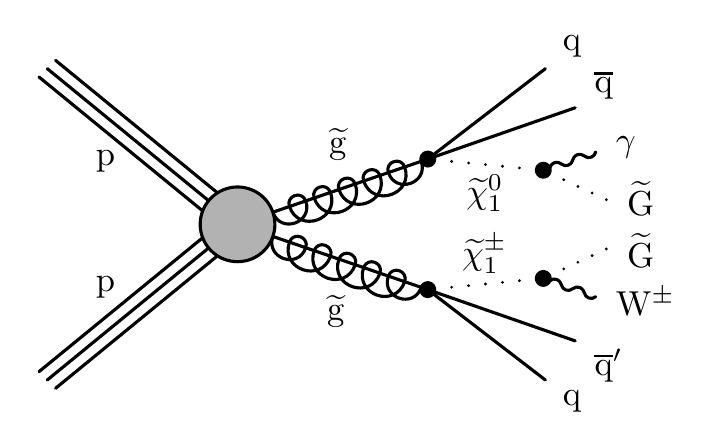}\\
\null
\includegraphics[width=\cmsFigWidth]{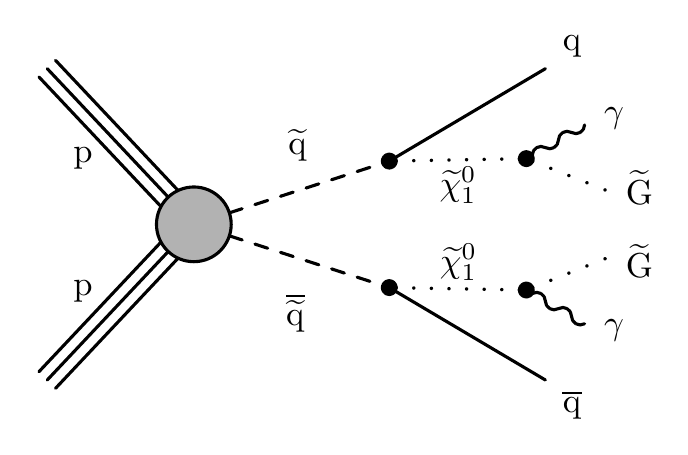}\hfil
\includegraphics[width=\cmsFigWidth]{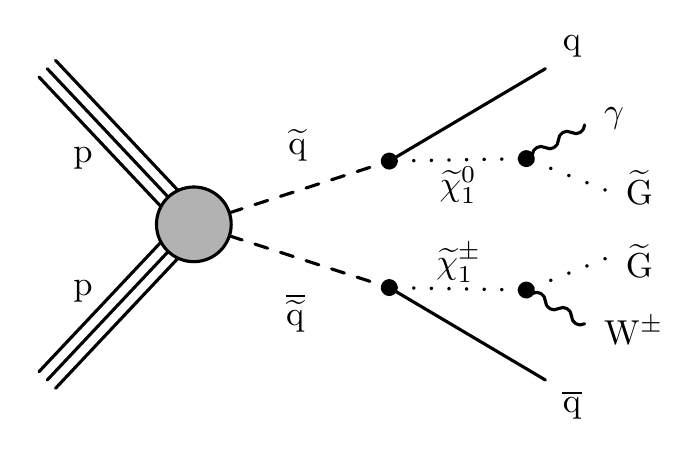}
\caption{
For strong gluino pair-production the simplified scenarios T5gg (upper left) and T5Wg (upper right) and for squark
pair-production the simplified scenarios T6gg (lower left) and T6Wg (lower right) are studied. In the T5Wg (T6Wg) scenario, a branching fraction of 50\% is assumed for the decays $\tilde{\text{g}}\to \PQq\PAQq\chipm_1$ and $\tilde{\text{g}}\to \PQq\PAQq\chiz_1$ ($\PSQ\to \PQq\chipm_1$ and $\PSQ\to \PQq\chiz_1$), resulting in final states with zero, one, or two photons.
\label{fig:feyn_strong}}
\end{figure*}

The strong production SMS models T5gg, T5Wg, T6gg, and T6Wg are shown in Fig.~\ref{fig:feyn_strong}, where T5gg and T5Wg represent gluino pair production, and T6gg and T6Wg squark pair production.
The neutralino decays as $\chiz_1\to\gamma\sGra$, while the chargino decays as $\chipm_1\to \PW^{\pm}\sGra$.
In the T5Wg and T6Wg scenario, a branching fraction of 50\% is assumed for the charged and neutral decays of the gluino or squark.
The T5gg (T6gg) scenario assumes a branching fraction of 100\% for $\tilde{\text{g}}\to \PQq\PAQq\chiz_1$ ($\PSQ\to \PQq\chiz_1$).

\section{The CMS detector}
\label{sec:detector}

The central feature of the CMS apparatus is a superconducting solenoid of 6\unit{m} internal dia\-meter, providing a magnetic field of 3.8\unit{T}. Within the solenoid volume are a silicon pixel and strip tracker, a lead tungstate crystal electromagnetic calorimeter (ECAL), and a brass and scintillator hadron calorimeter (HCAL), each composed of a barrel and two endcap sections. Extensive forward calorimetry complements the coverage provided by the barrel and endcap detectors. Muons are measured in gas-ionization detectors embedded in the steel flux-return yoke outside the solenoid.

In the barrel section of the ECAL, an energy resolution of approximately 1\% is achieved for unconverted or late-converting photons arising from the $\PH\to\Pgg\Pgg$ decay for photons with $\pt>25\GeV$. The remaining barrel photons have an energy resolution of about 1.3\% up to a pseudorapidity of $|\eta|=1$, rising to about 2.5\% at $|\eta|=1.4$. In the endcaps, the energy resolution of unconverted or late-converting photons is about 2.5\%, while the remaining endcap photons have a resolution between 3 and 4\%~\cite{Khachatryan:2015iwa}.

A more detailed description of the CMS detector, together with a definition of the coordinate system used and the relevant kinematic variables, can be found in Ref.~\cite{Chatrchyan:2008zzk}.

\section{Object reconstruction and simulation}
\label{sec:reconstruction}

The particle-flow (PF) event algorithm~\cite{Sirunyan:2017ulk} reconstructs and identifies each individual particle with an optimized combination of information from the various elements of the CMS detector. The energy of photons is directly obtained from the ECAL measurement, corrected for zero-suppression effects. Fully reconstructed photon conversions are used by the PF algorithm and are included in the set of photon candidates. The energy of electrons is determined from a combination of the electron momentum at the primary interaction vertex as determined by the tracker, the energy of the corresponding ECAL cluster, and the energy sum of all bremsstrahlung photons spatially compatible with originating from the electron track. The energy of muons is obtained from the curvature of the corresponding track. The energy of charged hadrons is determined from a combination of their momentum measured in the tracker and the matching ECAL and HCAL energy deposits, corrected for zero-suppression effects and for the response function of the calorimeters to hadronic showers. Finally, the energy of neutral hadrons is obtained from the corresponding corrected ECAL and HCAL energy.

Photons are reconstructed~\cite{Khachatryan:2015iwa} from clusters in the ECAL and are required to be isolated. The energy deposit in the HCAL tower closest to the seed of the ECAL supercluster~\cite{Khachatryan:2015hwa} assigned to the photon is required to be less than 5\% of the energy deposited in the ECAL. A photon-like transverse ECAL shower shape is required. The photon isolation is determined by computing the transverse energy in a cone centered around the photon momentum vector. The cone has an outer radius of 0.3 in $\Delta R = \sqrt{\smash[b]{(\Delta\phi)^2 + (\Delta\eta)^2}}$, where $\phi$ is the azimuthal angle, and the contribution of the photon is removed. Corrections for the effects of multiple interactions in the same or adjacent bunch crossing (pileup) are applied to all isolation energies, depending on the $\eta$ of the photon. To ensure that no photon with anomalously high a posteriori corrections populate the signal region, a requirement that at least 30\% of the photon's energy be deposited in the seed crystal is imposed for all considered photons. A photon candidate must exceed a minimal \pt of 15\GeV. Photons are efficiently discriminated against electrons by requiring that photons have no matching pattern of energy deposits in the pixel detector.

The vector \ptvecmiss is defined as the projection onto the plane perpendicular to the beams of the negative vector sum of the momenta of all PF candidates in an event. The magnitude of \ptvecmiss is referred to as \ptm.

Jets are reconstructed from PF candidates with the anti-\kt clustering algorithm~\cite{Cacciari:2008gp} as implemented in the {\sc FastJet}~\cite{fastjet} package, using a distance parameter of 0.4. Jet energy corrections~\cite{Cacciari:2007fd,2011JInst...611002C} are derived from Monte Carlo (MC) simulation, and are confirmed with in situ measurements of the energy balance in dijet and $\gamma$+jet events. These corrections are also propagated to \ptvecmiss. Jets with $\pt>30\GeV$ and $|\eta|<3$ are required to be geometrically isolated from identified photons, electrons, and muons, where electrons and muons have to fulfill standard identification requirements to be considered in this isolation criterion. Filters against anomalously high \ptm from instrumental effects are applied~\cite{Khachatryan:2014gga}.

The reconstructed vertex with the largest value of summed physics-object $\pt^2$ is taken to be the primary $\Pp\Pp$ interaction vertex. The physics objects are the jets, clustered using the jet finding algorithm~\cite{Cacciari:2008gp,fastjet} with the tracks assigned to the vertex as inputs, and the associated missing transverse momentum, taken as the negative vector sum of the \pt of those jets.

The SM background processes contributing to the signal and control regions are modeled using MC simulations. The quantum chromodynamics (QCD) multijet, \gammaJets, and $\PW$ and $\PZ$ processes are generated with \MGvATNLO 2.3.3~\cite{Alwall:2014hca,Alwall:2007fs} at leading order (LO), while the \ttgamma processes are generated at next-to-leading order (NLO)~\cite{Alwall:2014hca,Frederix:2012ps}. The $\PW\PW$ diboson production is generated with \textsc{Powheg} v2~\cite{Nason:2004rx,Frixione:2007vw,Alioli:2010xd,Melia:2011tj,Nason:2013ydw}, and $\PW\PZ$ and $\PZ\PZ$ production are generated using \PYTHIA{8}.205~\cite{Sjostrand:2006za}. The $\PZ\gamma$ sample is scaled with photon \pt dependent next-to-next-to-leading logarithmic (NNLL) K-factors~\cite{Bozzi:2010xn}, which are of the order of 1.3. A constant next-to-NLO (NNLO) K-factor of 1.34 is applied to the $\PW\gamma$ production cross section~\cite{Bozzi:2010xn}, and NLO K-factors of the order of 1.2 are applied to the $\PW$ and $\PZ(\to\nu\nu)$ production cross sections. The diboson production cross sections are available at NLO ($\PZ\PZ$, $\PW\PZ$) and NNLO ($\PW\PW$) precision~\cite{Gehrmann:2014fva}. The $\PW\gamma$ and $\PZ\gamma$ processes, collectively denoted as $\PV\gamma$, are the dominant backgrounds in the signal region. A data sideband region is used to obtain additional scale factors for the \Vgamma and \gammaJets samples, where \Vgamma comprises the \PW and \PZ boson production, with and without photon radiation.

The GGM signal scan is generated with \PYTHIA{8}, while the SMS signal scans are generated with \MGvATNLO at LO.
The cross sections are calculated at NLO and NLO+NLL accuracy \cite{Beenakker:1996ch, Kulesza:2008jb, Kulesza:2009kq, Beenakker:2009ha, Beenakker:2011fu, Borschensky:2014cia, Beenakker:1999xh, Fuks:2012qx, Fuks:2013vua} for the GGM and the SMS scans, respectively, with all the unconsidered sparticles assumed to be heavy and decoupled. For the EWK models, the cross sections are computed in a limit of mass-degenerate wino $\chiz_2$ and $\chipm_1$.

All MC samples incorporate the NNPDF 3.0~\cite{Ball:2014uwa} parton distribution functions (PDFs) and use the \PYTHIA v8.205 or \PYTHIA v8.212 program with the CUETP8M1 generator tune~\cite{Khachatryan:2015pea} to describe the parton showering and the hadronization.
Double counting of the partons generated with \MGvATNLO and those with \PYTHIA is removed using the MLM \cite{Alwall:2007fs} and the FXFX \cite{Frederix:2012ps} matching schemes, in the LO and NLO samples, respectively.
The \GEANTfour~\cite{Agostinelli:2002hh} package is used to model the detector and the detector response for SM processes, while the CMS fast simulation~\cite{1742-6596-331-3-032049,Sekmen:2017hzs} is used for signal samples. Additional $\Pp\Pp$ interactions are considered in the simulation and all samples are weighted on an event-by-event basis to match the distribution of the number of interaction vertices observed in data.

\section{Event selection}
\label{sec:selection}

The data are recorded using a trigger requiring one photon that passes very loose identification criteria and has a \pt of at least 165\GeV~\cite{CMS-PAPERS-TRG-12-001}. The events in the subsequent analysis are required to contain at least one identified and isolated photon with $\pt>180\GeV$ in the central barrel part of the detector ($\abs{\eta}<1.44$) that has been accepted by the trigger. The photons are required to have an angular distance in the $\eta{-}\phi$ plane of $\Delta R>0.5$ to the nearest jet.
To suppress events where the \ptm mainly arises from a significant mismeasurement of a jet's energy, all jets with $\pt>100\GeV$ must fulfill $\Delta\phi(\ptvecmiss,\text{jet})>0.3$, where $\Delta\phi(\ptvecmiss,\text{jet})$ is the distance in $\phi$ between the jet and the \ptm.
At least one reconstructed vertex per event is required~\cite{CMS-TDR-15-02}.
To maintain high signal acceptance for all studied signal scenarios no selection criteria are applied on the presence or absence of jets or leptons, except for the photon isolation criteria.
The photon trigger efficiency for this selection is found to be $\epsilon_\gamma=94.3\pm0.4\%$, independent of the kinematic event variables used in the analysis.

{\tolerance=900
The preselected events with at least one high-\pt photon are separated into a signal region and an orthogonal control region. The signal region is defined by $\ptm>300\GeV$ and $\mt>300\GeV$, where \mt is the transverse mass of the photon with the highest energy and \ptm, and roughly represents the NLSP mass in the SUSY scenarios containing the decay $\chiz_1\to\gamma\sGra$.
The requirement $\mt>300\GeV$ was chosen to optimize the statistics in the control region under maximization of the signal acceptances.
The region with $\ptm>100\GeV$ and $\mt>100\GeV$, but excluding the signal region, defines the signal-depleted data control region.
\par}

Multiple exclusive signal bins are defined with respect to $\ST\equiv\ptm+\sum_{\gamma_{i}}\pt(\gamma_i)$, the scalar sum of \ptm and the \pt of all photons in the event. The region with $\ptm>300\GeV$ and $\mt>300\GeV$, but $\ST\leq600\GeV$ has negligible signal contamination and is used to validate the background estimation. The four \ST regions 600--800, 800--1000, 1000--1300, and ${>}1300\GeV$ define exclusive bins that are simultaneously interpreted in a multichannel counting experiment for best sensitivity.
The full selection requirements to define each region used in this analysis are summarized in Table~\ref{tab:selections}.

\begin{table}[htb!]
\centering
\topcaption{Summary of the event selection criteria required for the control, validation, and signal regions. \label{tab:selections}}
\ifthenelse{\boolean{cms@external}}{\resizebox{\columnwidth}{!}}{}
{
\begin{tabular}{ll}
\multicolumn{1}{c}{Region}                            & \multicolumn{1}{c}{Selection} \\
\hline
& \ptm filters \\
& At least one reconstructed vertex\\
Preselection & At least one photon with $\pt > 180\GeV$\\
& $\Delta R(\gamma,\text{jet}) > 0.5$\\
& $\Delta\phi(\ptvecmiss,\text{jet}) > 0.3\unit{rad}$, if $\pt(\text{jet}) > 100\GeV$\\
\hline
& Preselection\\
Control  & $ \ptm > 100\GeV$\\
region & $\mt > 100\GeV$\\
& $\ptm < 300\GeV$ or $\mt < 300\GeV$\\
\hline
& Preselection\\
Validation  & $\ptm > 300\GeV$ \\
region & $\mt > 300\GeV$\\
& $\ST < 600\GeV$\\
\hline
& Preselection\\
Signal  & $\ptm > 300\GeV$\\
region & $\mt > 300\GeV$\\
& $\ST > 600\GeV$\\
\end{tabular}
}
\end{table}

The selection differs in several aspects from the analysis using 8\TeV data~\cite{SUS-14-016}. The trigger used in the 8\TeV analysis allowed for very low photon \pt and \ptm selections. The ``\ptm significance'' that defined the signal and control regions has been replaced by \ptm for simplicity and to allow for easier reinterpretations of the results. The analysis is optimized such that no loss in sensitivity is ensured.

\section{Background estimation}
\label{sec:analysis}

The SM background in the photon and \ptm final state is dominated by vector boson production with initial-state photon radiation, in particular by the $\Z\gamma\to\PGn\PAGn\gamma$ process. Direct photon production in association with jets, \gammaJets, also contributes at low values of \ptm and thus low values of \ST. A subdominant background arises from electrons misidentified as photons ($\Pe\to\gamma$). Further minor contributions originate from $\ttbar\gamma$ and diboson production. The most relevant backgrounds, \Vgamma and \gammaJets, are modeled by MC simulation and are scaled to the data in the data control region at low values of \ptm and \mt. The contribution from events with \efake misidentification is predicted from data. All remaining minor contributions are modeled by MC simulation.

\begin{figure}[tb]
\centering
\includegraphics[width=\cmsFigWidth]{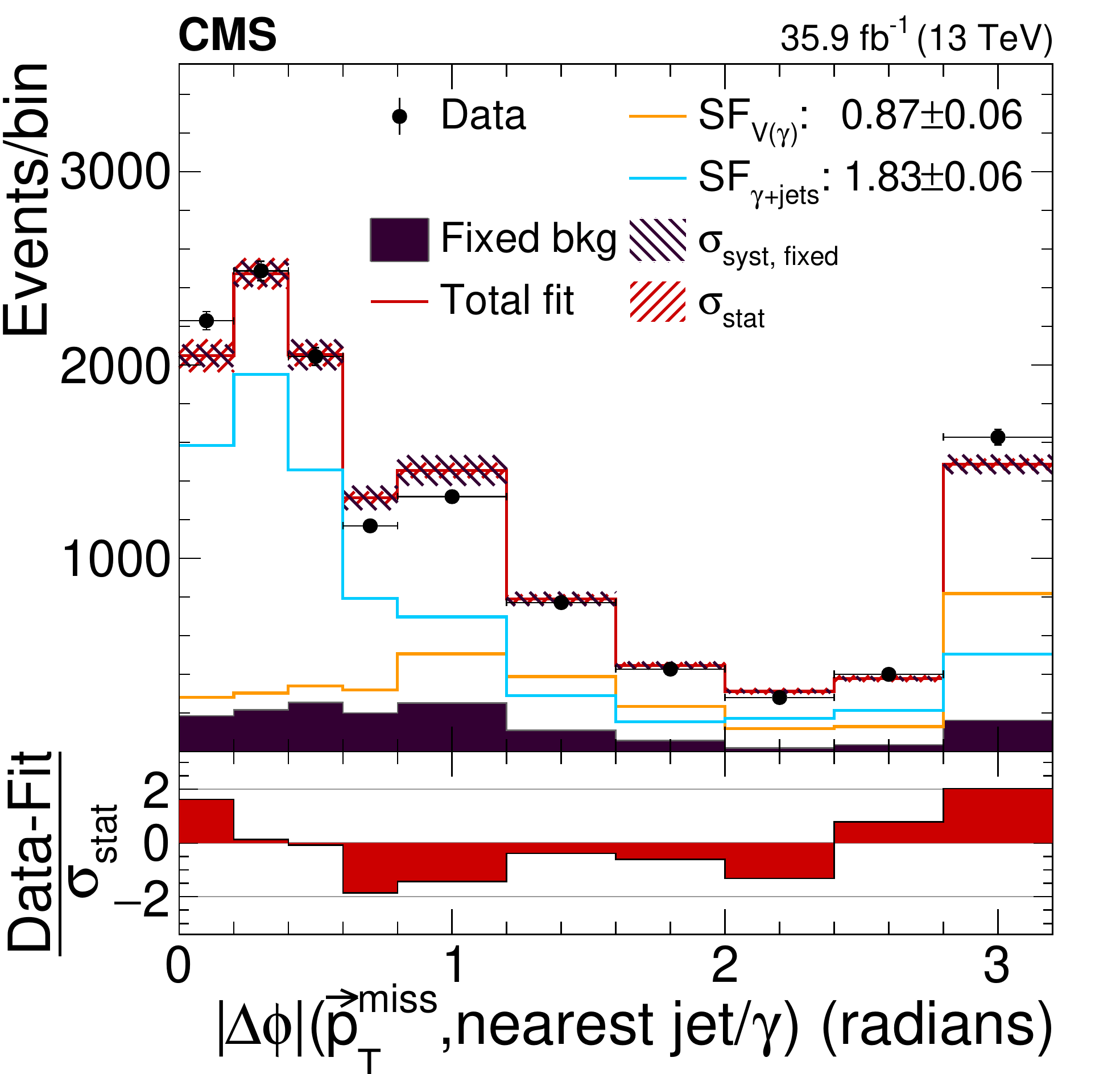}
\caption{The post-fit distributions for the \gammaJets (blue) and V($\gamma$) (orange) background in the control region
together with the fixed background (dark magenta) and the total fit distribution stacked onto the fixed backgrounds (red) are shown. The statistical uncertainty ($\sigma_{\text{stat}}$) of the post-fit distribution is shown in the red hatched area and the systematic uncertainty of the fixed background ($\sigma_{\text{syst, fixed}}$) is indicated with the dark magenta hatched area. The values SF$_{\Vgamma}$ and SF$_{\gammaJets}$ in the legend are the resulting scale factors. The pull distribution only considers the statistical uncertainty.
\label{fig:fit}}
\end{figure}

{\tolerance=4500
The normalization of the \Vgamma and \gammaJets backgrounds is determined in the control region by a simultaneous $\chi^2$-fit in bins of \dphi, which is the angular distance in the transverse plane of the \ptm and the nearest jet or photon.
The distribution of \dphi sufficiently separates the shapes of \Vgamma and \gammaJets backgrounds, so that scaling one background cannot compensate for the other. Contributions from other SM processes are small and are kept constant in the fit. Under the constraint of a fixed total yield, the scale factors for the \Vgamma and \gammaJets simulations are given by the minimum of the $\chi^2$ distribution. The resulting scale factors are
\begin{align}
\mathrm{SF}_{\Vgamma} &= 0.87\pm0.06, \\
\mathrm{SF}_{\gammaJets} &= 1.83\pm0.06,
\end{align}
where the uncertainties are of statistical origin only. The post-fit distribution of \dphi is shown in Fig.~\ref{fig:fit}. The size of the measured factors is consistent with the expectations~\cite{Bozzi:2010xn}. The scale factor for \Vgamma is smaller than unity because EWK corrections, which are not contained in the K-factors, are smaller than unity for high photon \pt. The \gammaJets scale factor is larger than unity since no K-factor is applied and QCD corrections for multijet backgrounds are large. The factors are found to be stable with respect to systematic variations of the method. Different control region selections, a variety of template variables, and various binnings of the template variables have been studied. Signal contamination becomes relevant if the gauginos are light because in terms of its kinematical variables the production of light gauginos is similar to that of \Vgamma production and is taken into account in the statistical analysis. In the remaining phase space, signal contamination is negligible.
\par}

Electrons that are misidentified as photons create a subdominant background, which can be predicted from data with good statistical precision. The misidentification rate $f_{\Pe\to\gamma}$ is measured in data in $\PZ\to\Pep\Pem$ decays with the ``tag-and-probe'' method~\cite{CMS:2011aa}. The dependence of the misidentification rate on the electron \pt and $\eta$ is studied. Nonresonant $\Pep\Pem$ background from non $\PZ$ boson events is estimated from $\Pe\mu$ events. The resulting misidentification rate in data is
\begin{equation}
f_{\Pe\to\gamma} = 2.7\pm1.3\%.
\end{equation}
The uncertainty of 50\% takes into account the variation of the misidentification rate as a function of the photon \pt, $\eta$, and several other variables.

\begin{figure*}[tb]
\centering
\includegraphics[width=\cmsFigWidth]{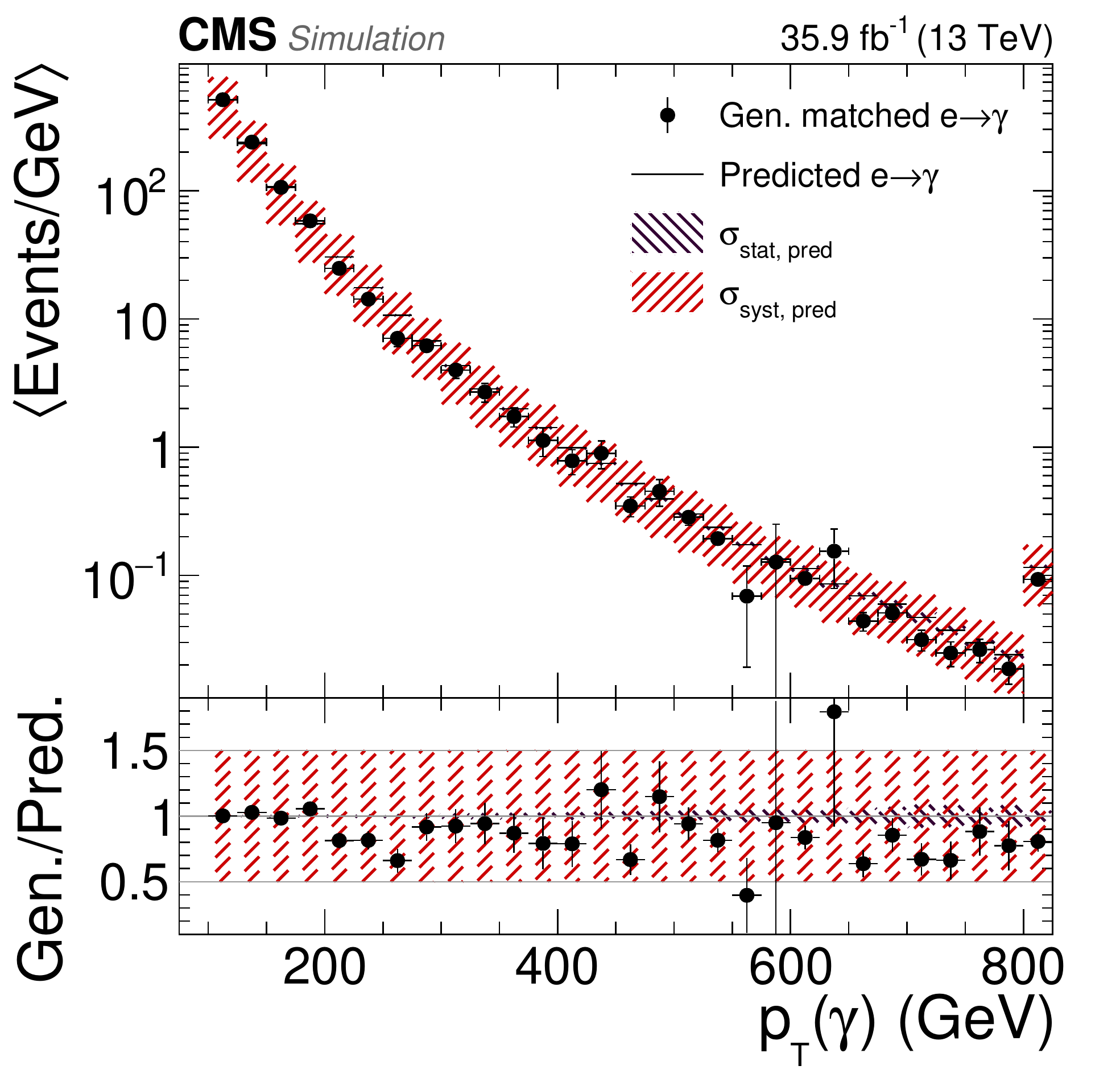}\hfil
\includegraphics[width=\cmsFigWidth]{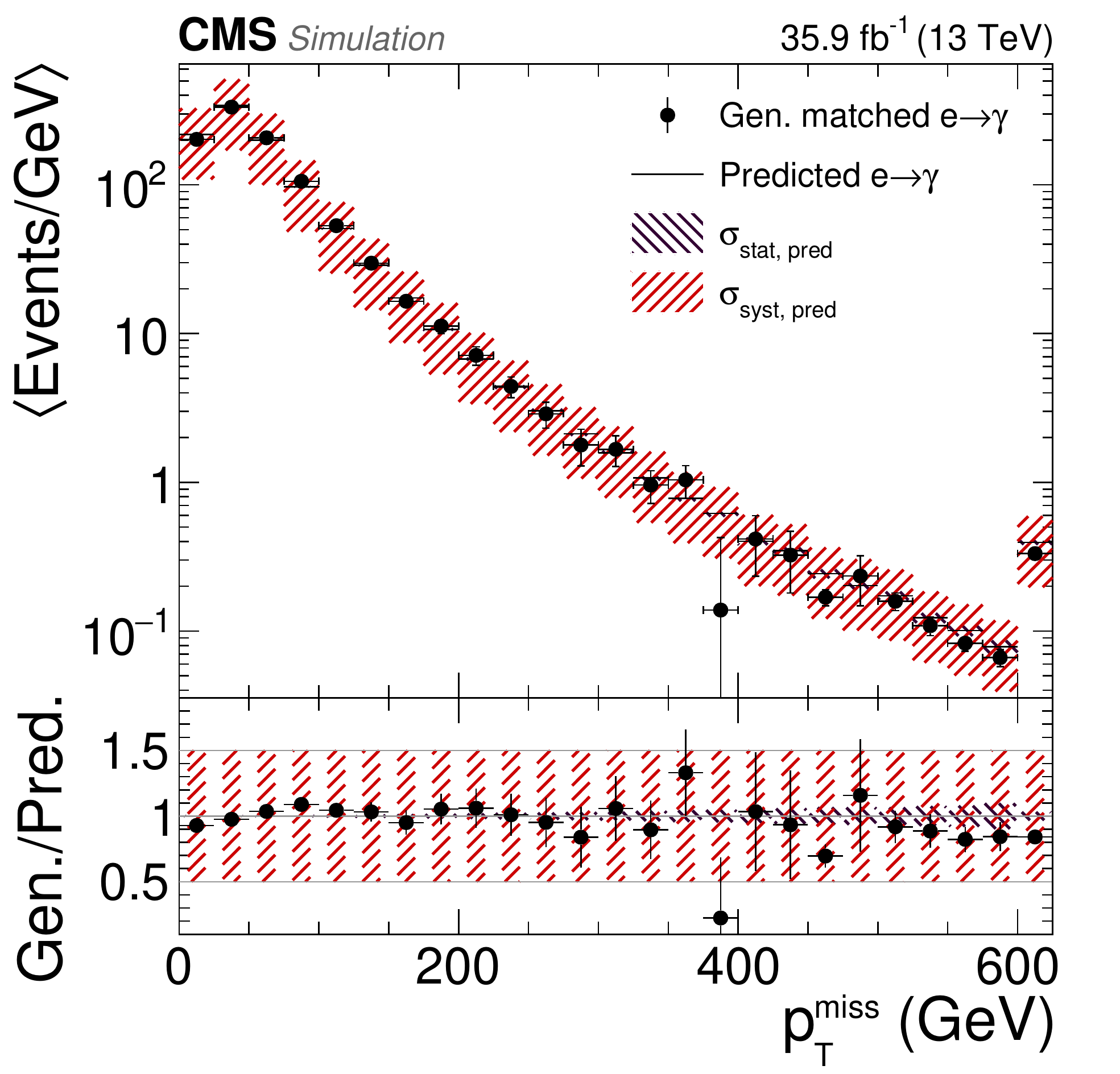}
\caption{Validation of the electron misidentification background estimation method using MC simulation. In the selection with at least one photon with $\pt>100$\GeV, the prediction of the $\Pe\to\gamma$ misidentification estimation method is compared to direct simulation in the photon \pt (left) and the \ptm (right) distributions. The black and red hatched areas represent the statistical ($\sigma_{\text{stat, pred}}$) and the 50\% systematic ($\sigma_{\text{syst, pred}}$) uncertainties of the prediction, respectively. Events populating the phase space beyond the shown range are included in the last bin.
\label{fig:egfake_closure}}
\end{figure*}

The $\Pe\to\gamma$ background is modeled from a data control sample with the same event selection as the signal region, but containing an identified electron instead of a photon. The sample is weighted by $f_{\Pe\to\gamma}$. The uncertainty of this estimation is dominated by the systematic uncertainties in the misidentification rate. The statistical uncertainty is negligible because the electron selection efficiency is about 40 times larger than $f_{\Pe\to\gamma}$. The method has been validated using MC simulation, as shown in Fig.~\ref{fig:egfake_closure}.

The minor contributions from $\ttbar(\gamma)$ and diboson processes are modeled using MC simulation as discussed above. Events where electrons are misidentified as photons are removed at the generator level to avoid overlaps. Based on simulation studies, the background from QCD multijet events is found to be negligible.

All uncertainties that would affect the normalization are eliminated for the \Vgamma and \gammaJets backgrounds by the MC normalization method. Therefore, the only remaining uncertainties originate from the simulated shape of these backgrounds.
The shape uncertainty due to the choice of the renormalization and factorization scales has been determined by varying these scales in different combinations of factors 0.5, 1, and 2 and repeating the fit of the \Vgamma and \gammaJets backgrounds. The prediction for each combination is compared in the four signal region bins for both backgrounds separately and bin-by-bin. The largest deviation in the respective bin is taken as the systematic uncertainty and varies in the range of 3.8--9.0\% and 2.8--7.1\% for the \Vgamma and \gammaJets backgrounds, respectively.
The LHC4PDF procedure~\cite{Butterworth:2015oua} is used to determine the shape uncertainty due to the choice of the PDFs and is determined bin-by-bin in the signal region and taken as systematic uncertainty, varying in the range of 1.6--3.8\% for the \Vgamma and 1.9--8.2\% for \gammaJets the background.
Although there is no direct usage of jets, the analysis is affected by the propagation of the jet energy scale (JES) uncertainty to \ptm. The resulting uncertainty affecting the final selection is determined by propagating the upward and downward shift of the JES to \ptm and repeating the analysis using the shifted \ptm. The largest deviation in the prediction is taken as systematic uncertainty and varies in the range of 5.0--5.9\% for the \Vgamma and 0.9--32\% for the \gammaJets background. The large deviation of 32\% for \gammaJets affects the highest bin in \ST, where only approximately one \gammaJets event is expected, so the absolute effect of this large uncertainty is small.
A 30\% uncertainty is assumed for the $\ttbar(\gamma)$ cross section, corresponding to a conservative estimate of the uncertainty with respect to the latest CMS measurement~\cite{Sirunyan:2017iyh}. The uncertainty in the diboson cross section is assumed to be 30\%.
Further systematic uncertainties, also affecting the signal simulation, arise from the trigger efficiency (0.4\%), the data to MC photon identification efficiency scale factor (2\%) and the
integrated luminosity (2.5\%)~\cite{CMS-PAS-LUM-17-001}.

\begin{table*}[bt!]
\centering
\topcaption{Systematic uncertainties in the background prediction in percent.
\label{tab:syst_unc}}
\begin{tabular}{lccccc}
& V($\gamma$)& $\gammaJets$ & $\Pe\to\Pgg$& $\ttbar(\Pgg)$&Diboson\\\hline
Fit uncert. of statistical origin&      6.9 &        $ 3.3$ &          \NA&             \NA&    \NA\\
Scale uncertainty in shape  &           3.8--9.0 &        2.8--7.1 &          \NA&             \NA&    \NA\\
PDF uncertainty in shape  &           1.6--3.8 &        1.9--8.2 &          \NA&             \NA&    \NA\\
JES uncertainty in shape  &           5.0--5.9 &        0.9--32\y &          \NA&             \NA&    \NA\\
Tag-and-probe fit                &          \NA&            \NA&      50 &             \NA&    \NA\\
Cross section, PDF, scales        &          \NA&            \NA&          \NA&         30 &30 \\
Integrated luminosity                       &          \NA&            \NA&          \NA&        2.5 &2.5\\
Photon eff. scale factor              &          \NA&            \NA&          \NA&          2.0 & 2.0  \\
Trigger efficiency               &          \NA&            \NA&          \NA&          0.4 & 0.4  \\
\end{tabular}
\end{table*}
\begin{table*}[tb!]
\centering
\topcaption{Systematic uncertainties in the signal predictions in percent.
\label{tab:signal_uncert}}
\begin{tabular}{lcc}
& \multicolumn{2}{c}{Signal scenario} \\
Source                            & EWK   & Strong production\\
\hline
Statistical MC precision per signal region &      $\x1{-}28$ &        $ \x2{-}50$  \\
Fast simulation uncertainty in \ptm	& ${<}0.1{-}5\x\y\z$ & ${<}0.1{-}25\y\z$ \\
Scale uncertainty in shape   &           ${<}0.1{-}1.8\z$ &        ${<}0.1{-}1.2\z$ \\
Integrated luminosity              &         2.5 &2.5\\
Trigger efficiency      &       0.4 & 0.4  \\
Photon scale factor	&		2.0 & 2.0\\
Pileup 		& ${<}$0.1--0.4\z & ${<}$0.1--2.1\z\\
ISR reweighting & 0.6--3.0 & \NA\\
\end{tabular}
\end{table*}

We improve the \MADGRAPH modeling of initial-state radiation (ISR), which affects the total transverse
momentum ($\pt^\text{ISR}$) of the system of SUSY particles, by reweighting the $\pt^\text{ISR}$ distribution of MC SUSY events. This reweighting procedure is based on studies of the \pt of $\PZ$ boson events~\cite{Chatrchyan:2013xna}. The reweighting factors range between 1.18 at $\pt^\text{ISR}=125\GeV$ and
0.78 for $\pt^\text{ISR}>600\GeV$. We take the deviation from unity as the systematic uncertainty in the reweighting procedure.

The systematic uncertainties affecting the background prediction and the signals are summarized in Tables~\ref{tab:syst_unc} and~\ref{tab:signal_uncert}, respectively.

\begin{figure*}[tb]
\centering
\includegraphics[width=\cmsFigWidth]{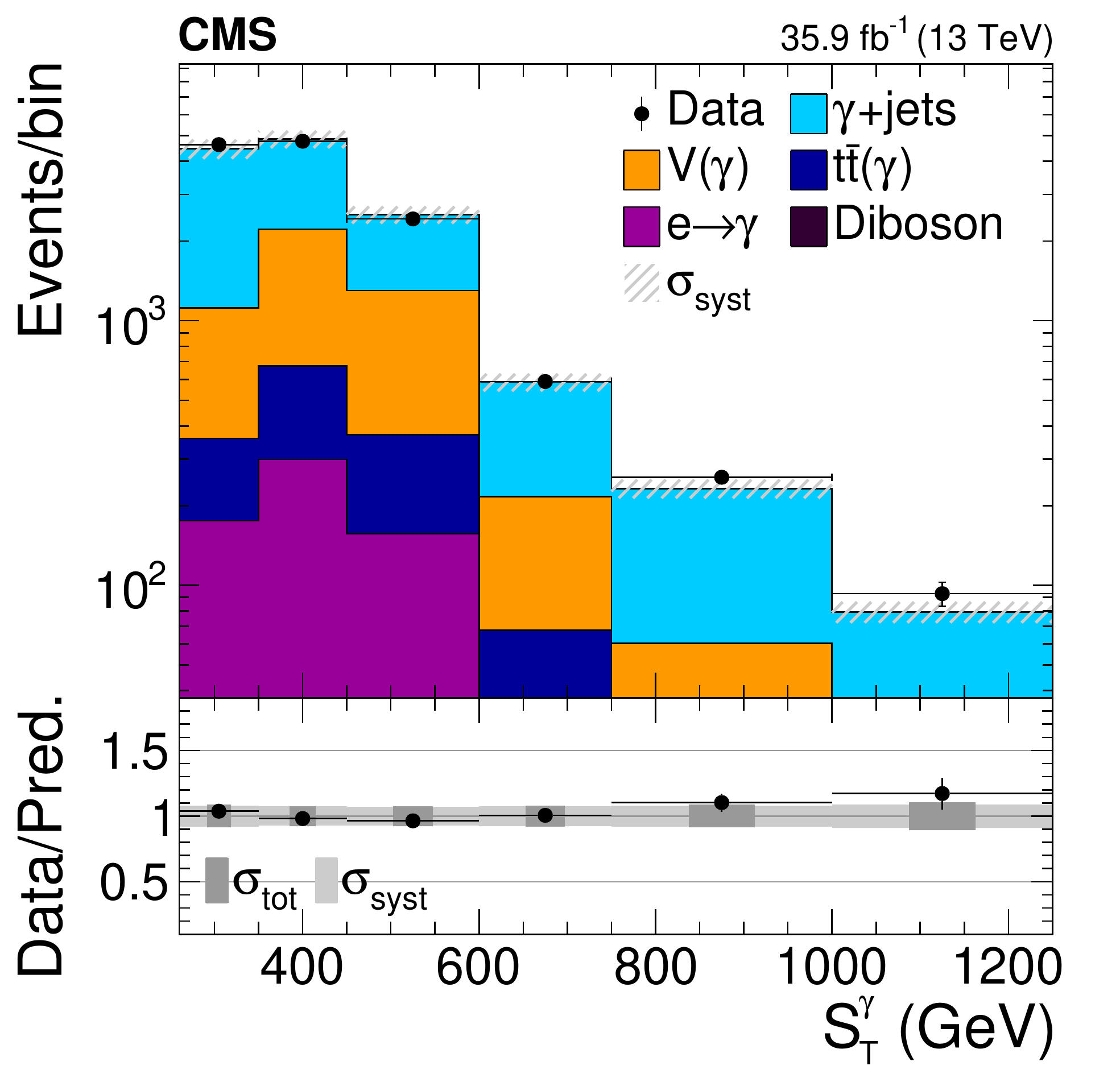}\hfil
\includegraphics[width=\cmsFigWidth]{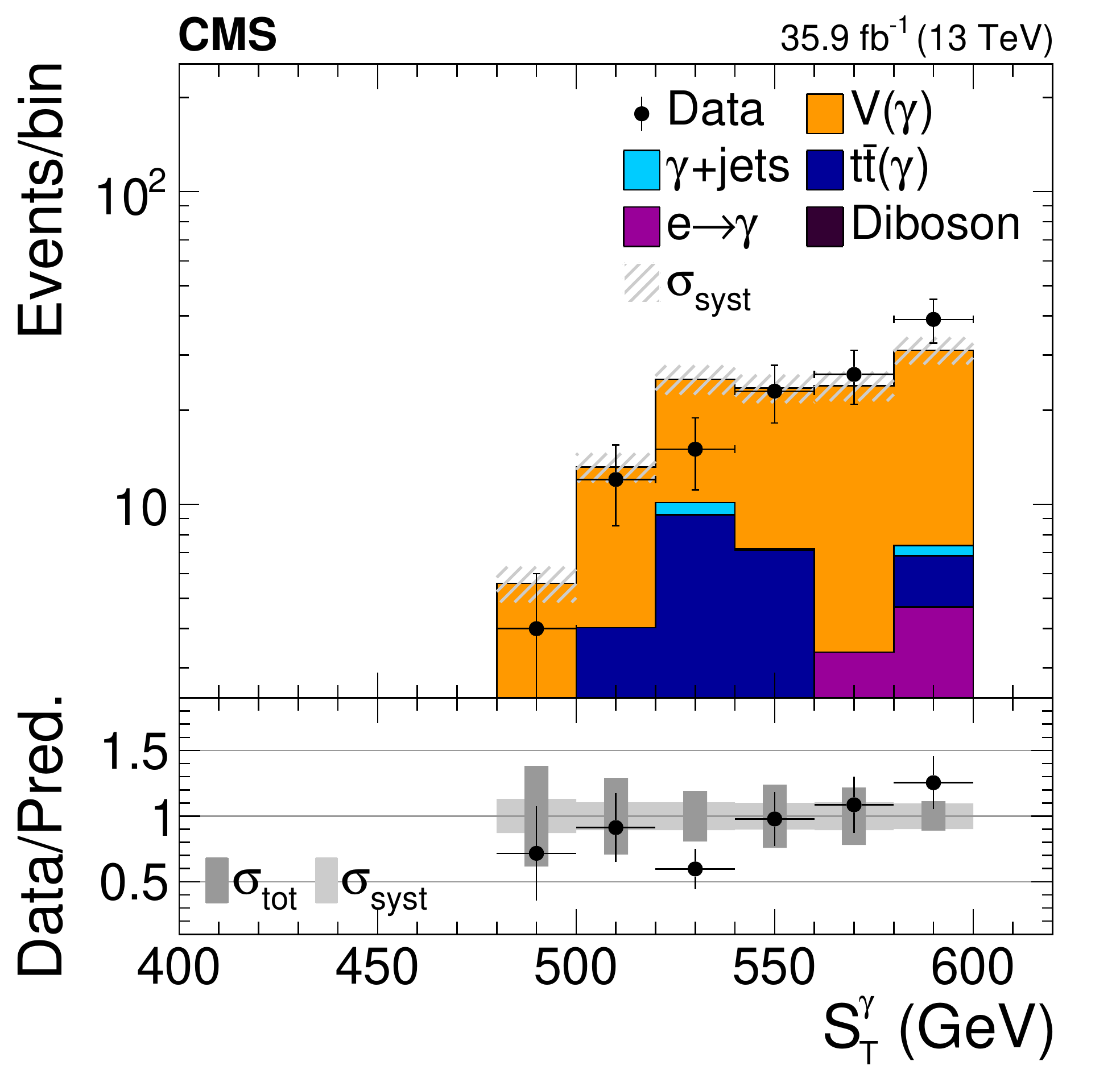}
\caption{Data to simulation comparisons in the control region (left) and the validation region (right). Events with \ST beyond the shown range are included in the last bin. The hatched light gray band in the upper panel, as well as the solid light gray band in the lower panel represent the total systematic uncertainty ($\sigma_{\text{syst}}$). The dark gray band in the lower panel indicates the quadratic sum of the statistical and systematic uncertainties ($\sigma_{\text{tot}}$).
\label{fig:data_mc}}
\end{figure*}

In Fig.~\ref{fig:data_mc} the signal sensitive variable \ST is shown for the control selection, used to derive scale factors for the \gammaJets and \Vgamma simulated samples, and for the validation selection. Good agreement is observed between the selected data and the SM background prediction.

\section{Results and interpretation}
\label{sec:results}

\begin{figure}[htb!]
\centering
\includegraphics[width=\cmsFigWidth]{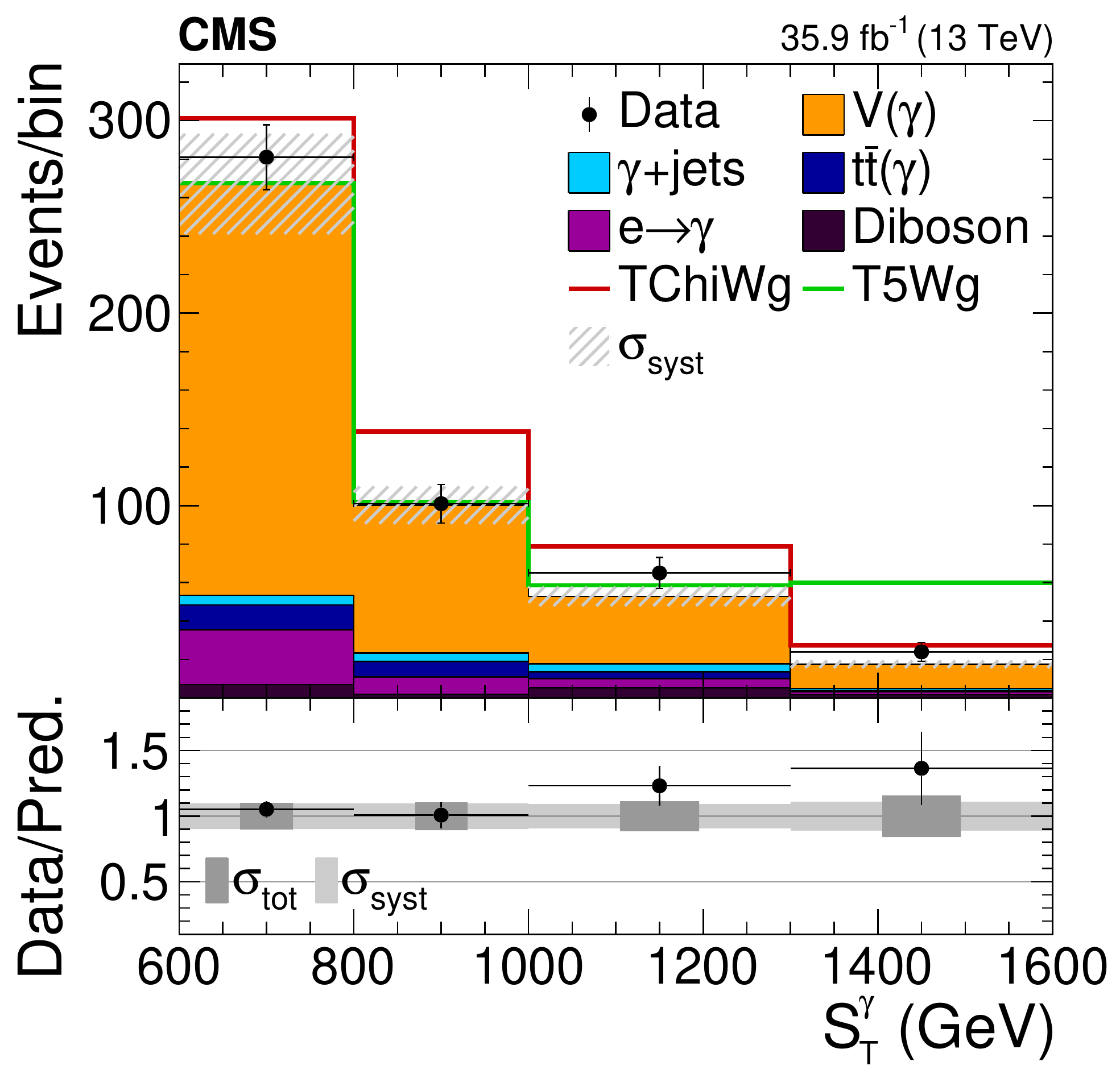}
\caption{Comparison of the measurement and prediction in the signal region in four exclusive bins of \ST. For guidance, two SUSY benchmark signal points are stacked on the SM background prediction, where the TChiWg signal point corresponds to a NLSP mass of 700\GeV and the T5Wg signal point corresponds to a gluino mass of 1750\GeV and a NLSP mass of 1700\GeV. Events with values of \ST beyond the shown range are included in the last bin. The hatched light gray band in the upper panel, as well as the solid light gray band in the lower panel represent the total systematic uncertainty ($\sigma_{\text{syst}}$). The dark gray band in the lower panel indicates the quadratic sum of the statistical and systematic uncertainties ($\sigma_{\text{tot}}$).
\label{fig:results}}
\end{figure}

Distributions of \ST in the four search regions are shown in Fig.~\ref{fig:results}. The corresponding yields are given in Table \ref{tab:sr_yields} for each bin, also showing the contributions of the individual background components. The statistical uncertainty in the \efake{} background is caused by the limited size of the collected data sample.
All other statistical uncertainties are due to the limited number of simulated events. The total systematic uncertainty results from the quadratic sum of the systematic uncertainties of each background component.
Good agreement is observed between the SM background prediction and the recorded data, without indication for the presence of new physics.

\begin{table*}[htb!]%
\centering
\topcaption{Background and data yields, as well as the statistical and systematic uncertainties for the separate signal region bins.
For the total background uncertainty the uncertainties of the individual background components are summed quadratically.
\label{tab:sr_yields}}
\begin{tabular}{lrcr}
\multicolumn{4}{c}{\ST{}: 600--800\GeV}\\
&  Yield & $\sigma_\text{stat}$ &  $\sigma_\text{syst}$ \\\hline
\Vgamma                    & 213 &   4.4 &  21.3 \\
\gammaJets                 &   5 &   1.1 &  0.5  \\
$\ttbar(\gamma)$          &  13 &   5.7 &   3.9\\
$\Pe\to\gamma$&  29 &   0.9 &  14.2 \\
Diboson                    &   7 &   2.8 &   2.1\\
\hline
Total                      & 267 &   7.9 &  26.0 \\
Data                       &    281 &  &  \\
\end{tabular}\qquad
\begin{tabular}{lrcc}
\multicolumn{4}{c}{\ST{}: 800--1000\GeV}\\
&  Yield & $\sigma_\text{stat}$ &  $\sigma_\text{syst}$ \\\hline
\Vgamma                    & 76.8 &   1.9 &  8.1 \\
\gammaJets                 &   4.4 &   1.2 &   0.4 \\
$\ttbar(\gamma)$          &  8.0 &   3.8 &   2.4\\
$\Pe\to\gamma$&  9.2 &   0.5 &  4.6 \\
Diboson                    &   1.9 &   1.7 &   0.6\\
\hline
Total                      & 100.2 &   4.7 &  9.7 \\
Data                       &    101\y\x &  &  \\
\end{tabular}
\\[4ex]
\begin{tabular}{lrcc}
\multicolumn{4}{c}{\ST{}: 1000--1300\GeV}\\
&  Yield & $\sigma_\text{stat}$ &  $\sigma_\text{syst}$ \\\hline
\Vgamma                       & 35.0 &   1.3 &  3.9 \\
\gammaJets                    &   4.2 &   1.3 &   0.4 \\
$\ttbar(\gamma)$             &  3.5 &   0.9 &   1.1\\
$\Pe\to\gamma$   &  4.7 &   0.4 &  2.3 \\
Diboson                       &   5.4 &   3.0 &   1.6\\
\hline
Total                         & 52.8 &   3.6 &  5.0 \\
Data                          &    65\y\x &  &  \\
\end{tabular}\qquad
\begin{tabular}{lrcc}
\multicolumn{4}{c}{\ST{}: ${>}$1300\GeV}\\
&  Yield & $\sigma_\text{stat}$ &  $\sigma_\text{syst}$ \\\hline
\Vgamma                       & 12.6 &   0.7 &  1.6 \\
\gammaJets                    &   1.1 &   0.5 &   0.4 \\
$\ttbar(\gamma)$             &  0.7 &   0.5 &   0.2\\
$\Pe\to\gamma$   &  1.5 &   0.2 &  0.8 \\
Diboson                       &   1.7 &   1.7 &   0.5\\
\hline
Total                         & 17.6 &   2.0 &  1.9 \\
Data                          &    24\y\x &  &  \\
\end{tabular}
\end{table*}

\begin{figure}[htb!]
\centering
\includegraphics[width=\cmsFigWidth]{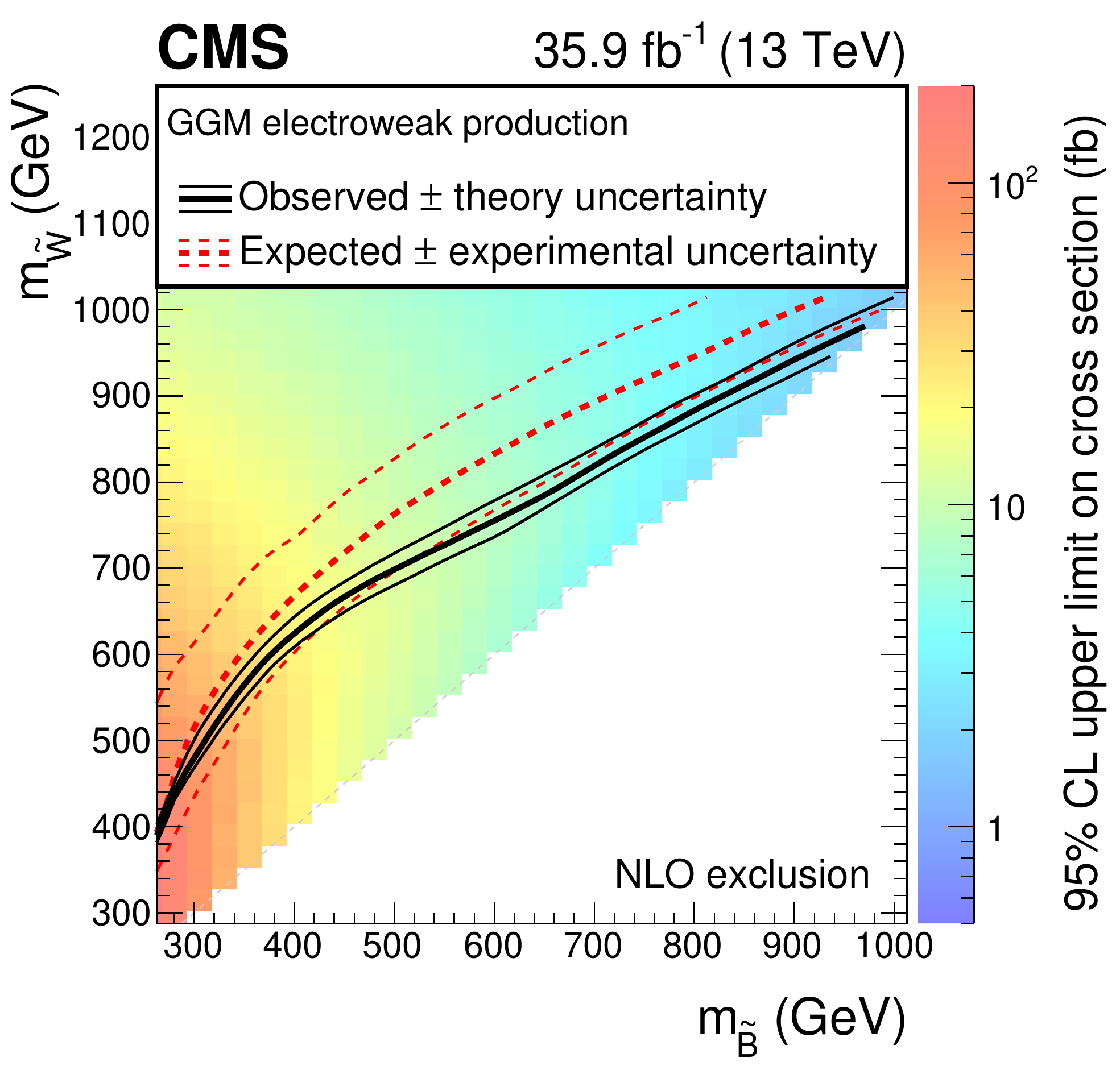}
\caption{Observed upper cross section limit at 95\%~CL for the EWK GGM signal in the wino-bino mass plane. The thick lines represent the observed (black) and expected (red) exclusion contours, where the phase space closer to the diagonal is excluded by the analysis. The thin dotted red curves indicate the region containing 68\% of the distribution of limits expected under the background-only hypothesis. The thin solid black curves show the change in the observed limit due to variation of the signal cross sections within their theoretical uncertainties.
\label{fig:limit_GGM}}
\end{figure}

Limits are calculated in one- and two-dimensional parameter spaces for the EWK and strong production models introduced in Section~\ref{sec:introduction}.
Upper limits on the signal cross section are calculated at 95\% confidence level (CL) using a modified frequentist CL$_\mathrm{s}$ approach~\cite{Junk:1999kv,Read,LHCCLs}
with a profile likelihood test statistic and asymptotic formulae~\cite{Cowan:2010js}.
The 95\% CL observed upper cross section limit, as well as the expected and observed exclusion contours, for the EWK GGM signal scan are shown in Fig.~\ref{fig:limit_GGM}. The limits are presented in the wino-bino mass plane. The analysis reaches the highest sensitivity for nearly degenerate wino and bino masses. In this case, the analysis excludes wino and bino masses up to 980\GeV at 95\%~CL, improving on the former best limit of 710\GeV~\cite{SUS-14-016}. The sensitivity decreases with a larger wino-bino mass splitting since on average the energy of the photons and gravitinos decreases, while more energy is transfered to the other decay products of the $\chipm_1$ and $\chiz_2$.

\begin{figure*}[htb!]
\centering
\includegraphics[width=\cmsFigWidth]{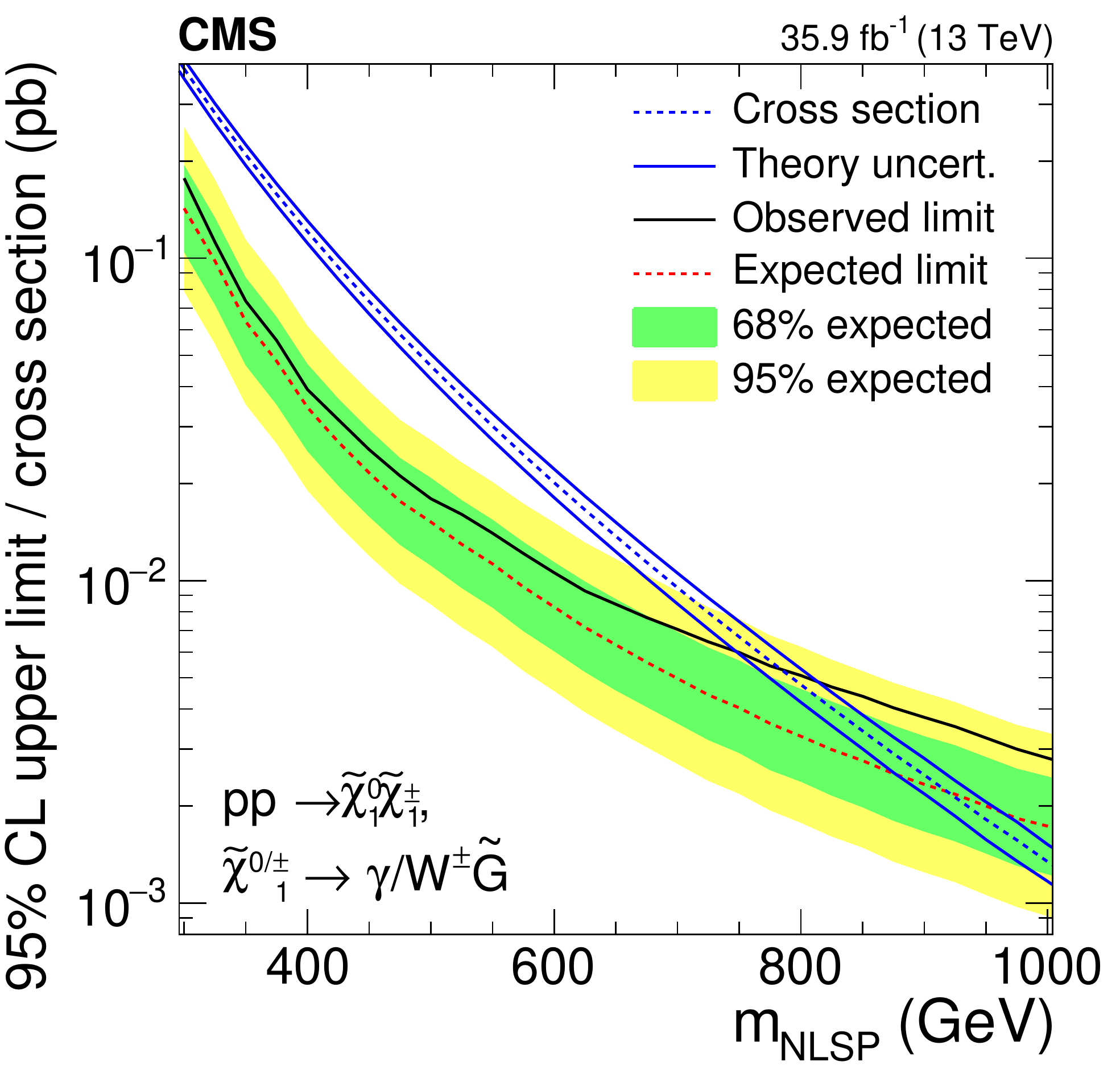}\hfil
\includegraphics[width=\cmsFigWidth]{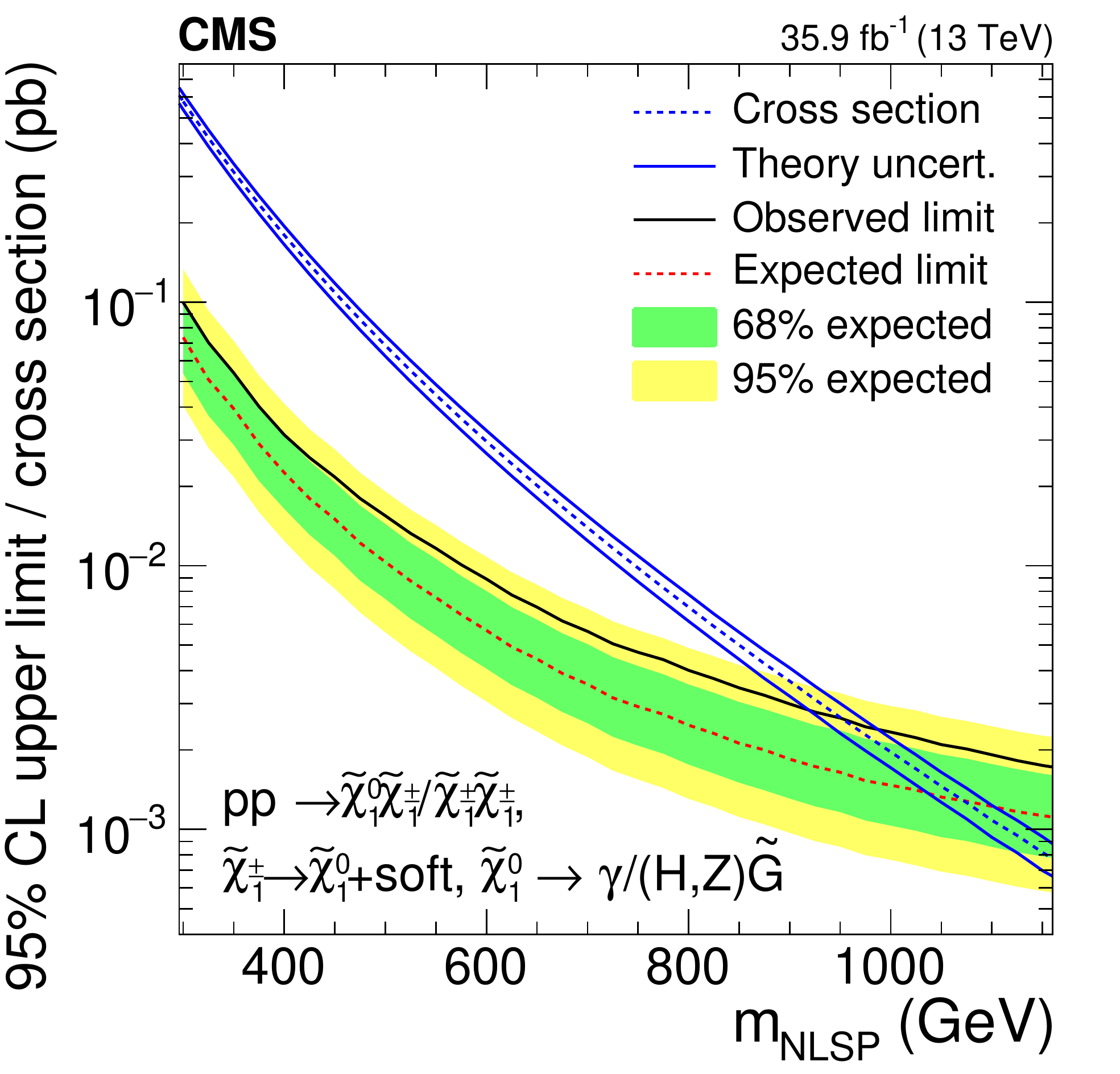}
\caption{Observed (black) and expected (red) upper cross section limits as a function
of the NLSP mass for the TChiWg (left) and TChiNg (right) model together with the corresponding theoretical cross section (blue). The inner (green) band and the outer (yellow) band indicate the regions containing 68 and 95\%, respectively, of the distribution of limits expected under the background-only hypothesis. The solid blue lines represent the theoretical uncertainty in the signal cross section.
\label{fig:limit_TChiWg}}
\end{figure*}

The limits for the EWK TChiWg and TChiNg simplified models are shown as a function of $m_\text{NLSP}$ in Fig.~\ref{fig:limit_TChiWg}
together with the theoretical cross section.
The analysis excludes NLSP masses below 780\GeV at 95\% CL in the TChiWg scenario and below 950\GeV in the TChiNg scenario. Due to the slight excess observed with respect to the SM background prediction especially in the highest \ST bins, the observed limits are weaker than the expected exclusion limits of 920\,(1070)\GeV for the TChiWg\,(TChiNg) scenario.

\begin{figure*}[htb!]
\centering
\includegraphics[width=\cmsFigWidth]{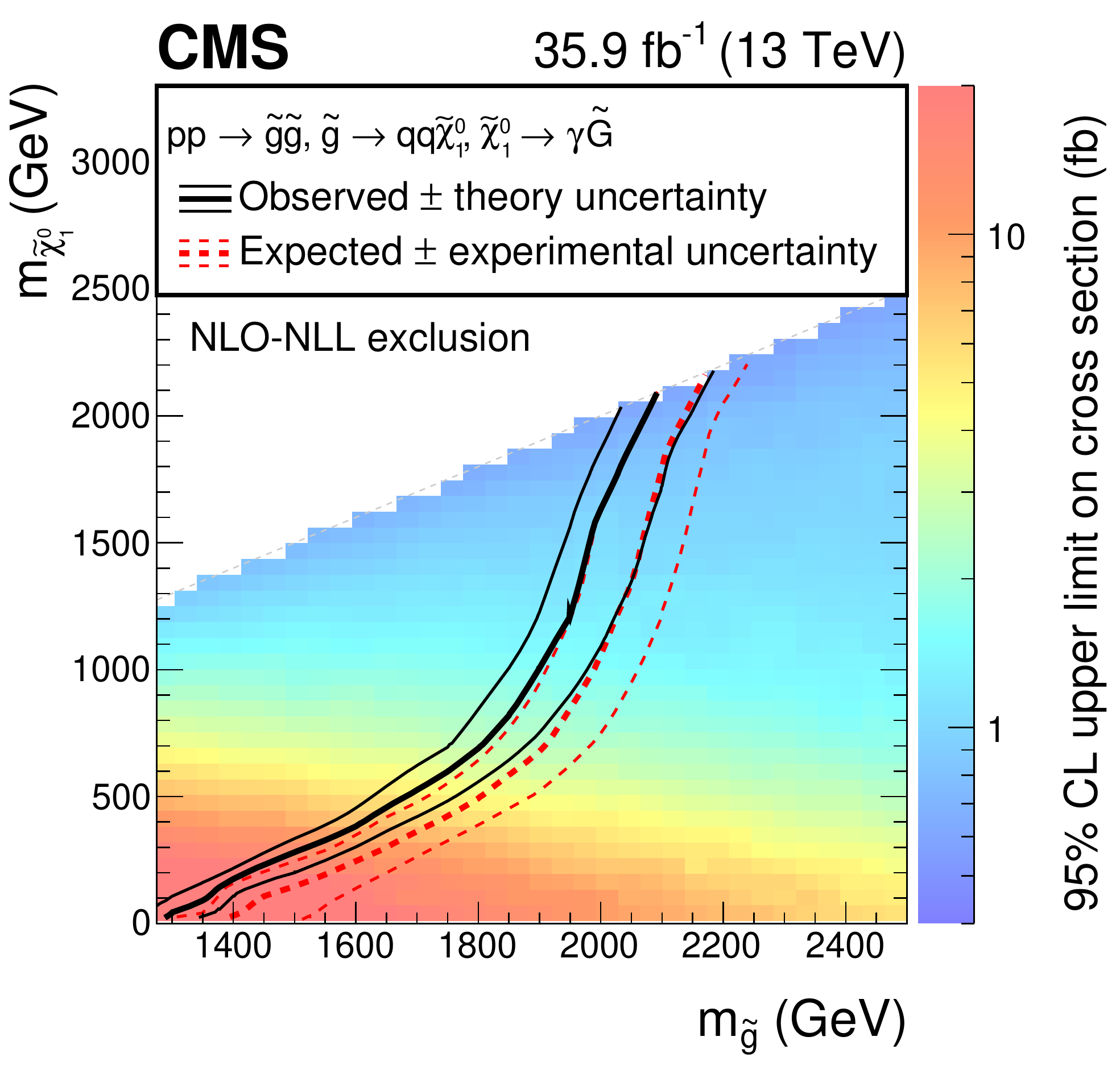}\hfil
\includegraphics[width=\cmsFigWidth]{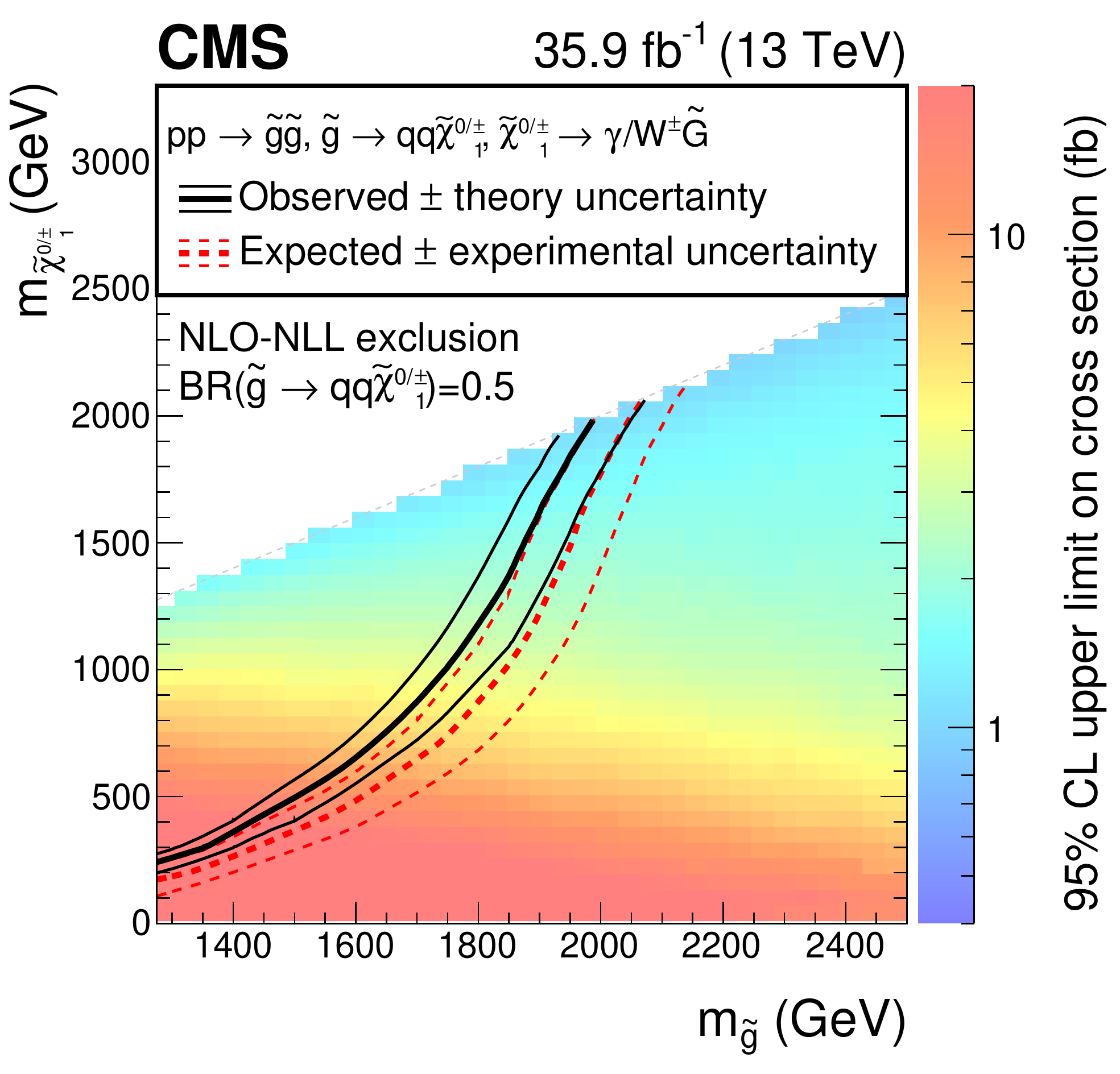}
\caption{The 95\% CL limits for the T5gg (left) and T5Wg (right) SMS models in the gluino-neutralino/chargino mass plane.
The color scale encodes the observed upper cross section limit for each point.
The thick lines represent the observed (black) and expected (red) exclusion contours, where the phase space of lower masses is excluded by the analysis. The thin dotted red curves indicate the region containing 68\% of the distribution of limits expected under the background-only hypothesis. The thin solid black curves show the change in the observed limit due to variation of the signal cross sections within their theoretical uncertainties.
\label{fig:limit_T5gg}}
\end{figure*}

The results are also interpreted in simplified models of strong production scenarios. The two scenarios T5gg and T5Wg represent the gluino pair production with two photons and one photon and one $\PW$ boson in the final state, respectively.
The cross section limits and exclusion contours are shown in Fig.~\ref{fig:limit_T5gg} in the $\PSg-\chiz_1/\chipm_1$ mass plane.
This search can exclude gluino masses of up to 2100\,(2000)\GeV in the T5gg\,(T5Wg) scenario.
The limit gets weaker at low NSLP masses because of the acceptance loss, which mostly arises from the lower energy of the photons and the gravitinos accompanied by larger hadronic activity in the event.

\begin{figure*}[htb!]
\centering
\includegraphics[width=\cmsFigWidth]{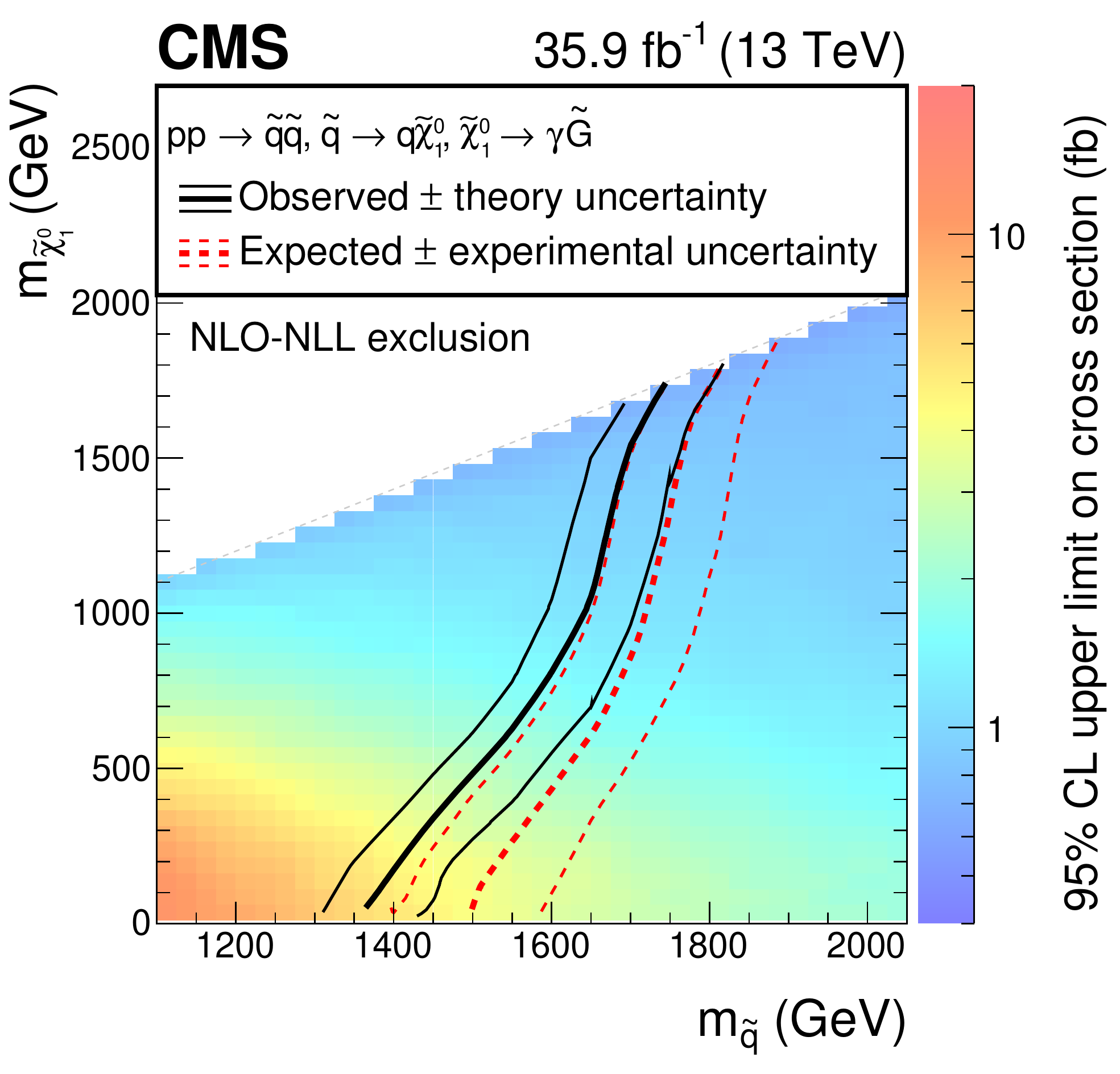}\hfil
\includegraphics[width=\cmsFigWidth]{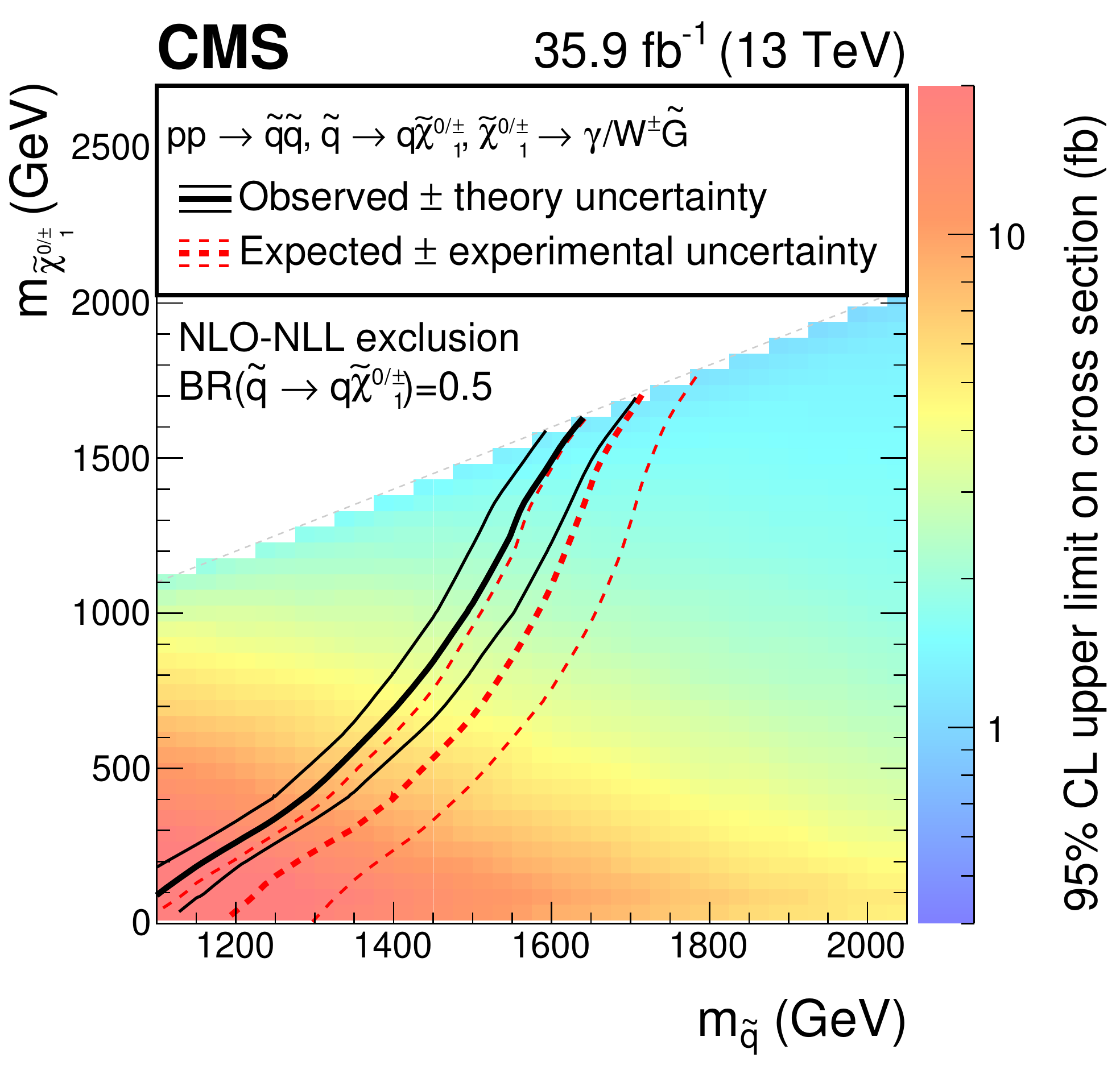}
\caption{The 95\% CL limits for the T6gg (left) and T6Wg (right) SMS models in the squark-neutralino/chargino mass plane.
The color scale encodes the observed upper cross section limit for each point.
The thick lines represent the observed (black) and expected (red) exclusion contours, where the phase space of lower masses is excluded by the analysis. The thin dotted red curves indicate the region containing 68\% of the distribution of limits expected under the background-only hypothesis. The thin solid black curves show the change in the observed limit due to variation of the signal cross sections within their theoretical uncertainties. For the signal production cross section five accessible mass-degenerate squark flavors for $\PSQ_{\text{L}}$ and $\PSQ_{\text{R}}$ were assumed.
\label{fig:limit_T6gg}}
\end{figure*}

Similar scenarios, T6gg and T6Wg, based on squark production are also used for interpretation and are shown in Fig.~\ref{fig:limit_T6gg}. Here, squark masses up to 1750\,(1650)\GeV are excluded for T6gg\,(T6Wg).

The mass limits on squarks are weaker compared to those on gluinos due to the generally lower production cross section. However, for squark production the hadronic activity in the event is lower compared to gluino production, slightly reducing the dependence on the $\PSQ-\chiz_1/\chipm_1$ mass difference.
The higher sensitivity in the T5gg and T6gg models is due to two photons contributing to \ST, increasing the separation power between the signal and the SM background.

\section{Summary}
\label{sec:conclusion}

A search for electroweak (EWK) and strong production of gauginos in the framework of gauge mediated supersymmetry breaking in final states with photons and large missing transverse momentum has been performed. A data set recorded by the CMS experiment at a center-of-mass energy of 13\TeV, corresponding to an integrated luminosity of 35.9\fbinv, was analyzed. The data were found to agree with the expectation from the standard model, without any indication of new physics.

The analysis is sensitive to EWK production of gauginos and to strong production of gluinos and squarks in particular if the mass difference between gauginos and gluinos or squarks is small.
A two-dimensional EWK signal scan in the framework of general gauge mediation is used to interpret the results. In the case of similar wino and bino masses, the analysis excludes masses below 980\GeV at 95\% confidence level, improving on the current best limit by 270\GeV~\cite{SUS-14-016}.
Two EWK simplified models are also used for the interpretation. The analysis excludes masses of the next-to-lightest supersymmetric particle $\chiz_1$ below 780 (950)\GeV in the TChiWg (TChiNg) scenario.
Additionally, limits are set for strong production simplified models based on gluino (T5gg, T5Wg) and squark (T6gg, T6Wg) pair production, excluding gluino (squark) masses up to 2100 (1750)\GeV.
This analysis complements searches in the photon+jets, diphoton, and photon+leptons final states, and sets the most stringent limits to date in the EWK production models, and in the strong production models when the gauginos are degenerate in mass with the gluino or squarks.

\begin{acknowledgments}

\hyphenation{Bundes-ministerium Forschungs-gemeinschaft Forschungs-zentren Rachada-pisek} We congratulate our colleagues in the CERN accelerator departments for the excellent performance of the LHC and thank the technical and administrative staffs at CERN and at other CMS institutes for their contributions to the success of the CMS effort. In addition, we gratefully acknowledge the computing centres and personnel of the Worldwide LHC Computing Grid for delivering so effectively the computing infrastructure essential to our analyses. Finally, we acknowledge the enduring support for the construction and operation of the LHC and the CMS detector provided by the following funding agencies: BMWFW and FWF (Austria); FNRS and FWO (Belgium); CNPq, CAPES, FAPERJ, and FAPESP (Brazil); MES (Bulgaria); CERN; CAS, MoST, and NSFC (China); COLCIENCIAS (Colombia); MSES and CSF (Croatia); RPF (Cyprus); SENESCYT (Ecuador); MoER, ERC IUT, and ERDF (Estonia); Academy of Finland, MEC, and HIP (Finland); CEA and CNRS/IN2P3 (France); BMBF, DFG, and HGF (Germany); GSRT (Greece); OTKA and NIH (Hungary); DAE and DST (India); IPM (Iran); SFI (Ireland); INFN (Italy); MSIP and NRF (Republic of Korea); LAS (Lithuania); MOE and UM (Malaysia); BUAP, CINVESTAV, CONACYT, LNS, SEP, and UASLP-FAI (Mexico); MBIE (New Zealand); PAEC (Pakistan); MSHE and NSC (Poland); FCT (Portugal); JINR (Dubna); MON, RosAtom, RAS, RFBR and RAEP (Russia); MESTD (Serbia); SEIDI, CPAN, PCTI and FEDER (Spain); Swiss Funding Agencies (Switzerland); MST (Taipei); ThEPCenter, IPST, STAR, and NSTDA (Thailand); TUBITAK and TAEK (Turkey); NASU and SFFR (Ukraine); STFC (United Kingdom); DOE and NSF (USA).

\hyphenation{Rachada-pisek} Individuals have received support from the Marie-Curie programme and the European Research Council and Horizon 2020 Grant, contract No. 675440 (European Union); the Leventis Foundation; the A. P. Sloan Foundation; the Alexander von Humboldt Foundation; the Belgian Federal Science Policy Office; the Fonds pour la Formation \`a la Recherche dans l'Industrie et dans l'Agriculture (FRIA-Belgium); the Agentschap voor Innovatie door Wetenschap en Technologie (IWT-Belgium); the Ministry of Education, Youth and Sports (MEYS) of the Czech Republic; the Council of Science and Industrial Research, India; the HOMING PLUS programme of the Foundation for Polish Science, cofinanced from European Union, Regional Development Fund, the Mobility Plus programme of the Ministry of Science and Higher Education, the National Science Center (Poland), contracts Harmonia 2014/14/M/ST2/00428, Opus 2014/13/B/ST2/02543, 2014/15/B/ST2/03998, and 2015/19/B/ST2/02861, Sonata-bis 2012/07/E/ST2/01406; the National Priorities Research Program by Qatar National Research Fund; the Programa Severo Ochoa del Principado de Asturias; the Thalis and Aristeia programmes cofinanced by EU-ESF and the Greek NSRF; the Rachadapisek Sompot Fund for Postdoctoral Fellowship, Chulalongkorn University and the Chulalongkorn Academic into Its 2nd Century Project Advancement Project (Thailand); the Welch Foundation, contract C-1845; and the Weston Havens Foundation (USA).

\end{acknowledgments}

\bibliography{auto_generated}

\cleardoublepage \appendix\section{The CMS Collaboration \label{app:collab}}\begin{sloppypar}\hyphenpenalty=5000\widowpenalty=500\clubpenalty=5000\textbf{Yerevan Physics Institute,  Yerevan,  Armenia}\\*[0pt]
A.M.~Sirunyan, A.~Tumasyan
\vskip\cmsinstskip
\textbf{Institut f\"{u}r Hochenergiephysik,  Wien,  Austria}\\*[0pt]
W.~Adam, F.~Ambrogi, E.~Asilar, T.~Bergauer, J.~Brandstetter, E.~Brondolin, M.~Dragicevic, J.~Er\"{o}, M.~Flechl, M.~Friedl, R.~Fr\"{u}hwirth\cmsAuthorMark{1}, V.M.~Ghete, J.~Grossmann, J.~Hrubec, M.~Jeitler\cmsAuthorMark{1}, A.~K\"{o}nig, N.~Krammer, I.~Kr\"{a}tschmer, D.~Liko, T.~Madlener, I.~Mikulec, E.~Pree, N.~Rad, H.~Rohringer, J.~Schieck\cmsAuthorMark{1}, R.~Sch\"{o}fbeck, M.~Spanring, D.~Spitzbart, W.~Waltenberger, J.~Wittmann, C.-E.~Wulz\cmsAuthorMark{1}, M.~Zarucki
% ...
\vskip\cmsinstskip
% ...
1:~~Also at Vienna University of Technology, Vienna, Austria\\
% ...

\end{sloppypar}
\end{document}